\documentclass[aps, prd, superscriptaddress, nofootinbib, preprintnumbers]{revtex4}

\usepackage[dvipdfmx]{graphicx}
\usepackage{graphicx,bm,color}

\usepackage{amsmath,amssymb,bm}
\usepackage{float}
\usepackage{braket}
\usepackage{comment}
\usepackage{ascmac}
\usepackage{subfigure}
\usepackage[utf8]{inputenc}
\usepackage{appendix}
\usepackage{ulem}

\newcommand{\fulltoday}{\number\day\space \ifcase\month\or
    January\or February\or March\or April\or May\or June\or
    July\or August\or September\or October\or November\or December\fi
    \space\number\year}

 \makeatletter
    
    \@addtoreset{equation}{section}
  \makeatother

\allowdisplaybreaks

\begin{document}


\title{
Analysis on nonperturbative flavor violation at chiral crossover criticality in QCD}


\author{Mamiya Kawaguchi,}\thanks{{\tt kawaguchi@fudan.edu.cn}} 
\affiliation{Department of Physics and Center for Field Theory and Particle Physics, Fudan University, 220 Handan Road, 200433 Shanghai, China}
\author{Shinya Matsuzaki,}\thanks{{\tt synya@jlu.edu.cn}} 
\affiliation{Center for Theoretical Physics and College of Physics, Jilin University, Changchun, 130012, China}
\author{Akio Tomiya}\thanks{{\tt akio.tomiya@riken.jp}}
\affiliation{RIKEN BNL Research center, Brookhaven National Laboratory, Upton, NY, 11973, USA}




\begin{abstract}
We discuss the violation of quark-flavor symmetry 
at high temperatures, induced from nonperturbative 
thermal loop corrections and axial anomaly, 
based on a three-flavor linear-sigma model 
including an $SU(3)$ flavor violation induced by $U(1)_A$ anomaly which we call
an axial-anomaly induced-flavor breaking term.  
We employ a nonperturbative analysis following 
the Cornwall-Jackiw-Tomboulis formalism, and show that 
the model undergoes a chiral crossover 
with a pseudo-critical temperature, consistently with lattice observations. 
We find following features regarding the flavor breaking 
eminent around and above the pseudo-critical temperature: 
i) up-and down-quark condensates drop 
faster than the strange quark's toward the criticality, but still keep nonzero value 
even going far above the critical temperature; 
ii) the introduced anomaly-related flavor-breaking effect acts as a catalyzer toward the chiral restoration, 
and reduces the amount of flavor breaking in the up, down and strange quark condensates; 
iii) a dramatic deformation for the meson flavor mixing structure is observed, in which the anomaly-induced favor breaking is found to be almost irrelevant;  
iv) the meson spectroscopy gets corrected by the net nonperturbative flavor breaking effects, 
where the scalar meson mass hierarchy (inverse mass hierarchy) is significantly altered by the presence of the anomaly-related flavor breaking; 
v) the topological susceptibility significantly gets 
the contribution from the surviving strange quark condensate, 
which cannot be dictated by the chiral perturbation 
theory, and deviates from the dilute instanton gas prediction. There 
the anomaly-induced flavor breaking plays a role 
of the destructive interference for the net flavor violation, 
as in the flavor breaking in the quark condensates; 
vi) the $U(1)_A$ breaking signaled by nonzero topological susceptibility is enhanced 
by the nonperturbative strange quark condensate, which  may account for the tension in the effective restoration of the $U(1)_A$ symmetry
currently observed on lattices with two flavors 
and 2+1 flavors near the chiral limit.
Our founding critical natures can be checked 
in the future lattice simulations, and will give 
some implications to 
the thermal history of QCD axion. 
\end{abstract}

\maketitle
\flushbottom

\section{Introduction}
The QCD phase transition, involving the chiral symmetry restoration, is the major subject to understand the QCD vacuum structure present in the early Universe, 
and would also provide hints for astrophysical consequences related to the QCD thermal history, such as the QCD axion. 
Recent developments on lattice calculations on hot QCD,  
with $N_f=2+1$ flavors and physical quark masses, have revealed that the chiral phase transition is predicted as a crossover (called the chiral crossover), with the  pseudo-critical temperature estimated as $T_{pc}\sim 155\, {\rm MeV}$~\cite{Aoki:2006we,Borsanyi:2011bn,Ding:2015ona,Bazavov:2018mes,Ding:2020rtq}. 
Besides, $U(1)_A$ symmetry is intriguing also to study the hot-QCD phase structure. 
The $U(1)_A$ symmetry breaking may crucially contribute to the chiral phase transition and its phase transition order~\cite{Pisarski:1983ms}: although the $U(1)_A$ symmetry is broken by the quantum anomaly at low temperatures, above the chiral phase transition the fate of $U(1)_A$ symmetry is still mysterious, has not completely been understood yet, 
and left with a long-standing issue. 
The $U(1)_A$ symmetry is correlated with the nontrivial-topological vacuum structure, from which the CP violation arises due to strong interactions (strong CP violation) with the $\theta$ parameter. 
Thus the $U(1)_A$ anomaly at finite temperature, as well as 
the chiral symmetry restoration, is an important key 
to deeply understand the hot QCD vacuum including the topological 
structure.

The $\theta$-dependence of the QCD vacuum energy is captured 
in part by the topological susceptibility, which is defined as the curvature of the free energy of QCD with respect to the $\theta$ at the QCD vacuum.  
Note that the topological susceptibility is closely linked with the chiral symmetry breaking along with the $U(1)_A$ anomaly: 
the topological gluon configurations as the source of the axial anomaly can always be transferred by the axial rotation to current quark mass terms in the QCD generating functional, once the current quark masses are introduced in the theory. 
Then the topological susceptibility is given by the sum of quark condensates coupled with current quark masses 
and pseudoscalar susceptibilities (as will more explicitly be clarified in the later section).
Thereby, the topological susceptibility at high 
temperatures would be a crucial 
quantity to explore 
some (effective or partial) $U(1)_A$ restoration in the presence of nonzero current quark masses, in relation 
to the hot-QCD $\theta$ vacuum structure, and 
the chiral symmetry restoration via quark condensates.

In the three flavor QCD
at the vacuum the quark condensates are degenerated ($\langle \bar{u}u \rangle \simeq \langle \bar{d}d \rangle \simeq \langle \bar{s}s \rangle$) as observed in the lattice simulation, $\langle \bar{s}s \rangle/\langle \bar{l}l \rangle = 1.08 \pm 0.16$ ($l=u,d$)~\cite{McNeile:2012xh}.
Turning finite temperature on, this situation is dramatically altered. 
As has been observed in several analyses on lattice simulations for finite tempeture QCD~\cite{Cheng:2007jq,Nicola:2016jlj}, thermal loop effects would cause a partial restoration of the chiral symmetry (or chiral crossover) at the (pseudo) critical temperature, where only the lightest $l$ quark condensate $\langle \bar{l} l \rangle$ drops more quickly than the strange quark $\langle \bar{s}s \rangle$ does. 
This implies a nonperturbative flavor breaking generated in hot-QCD. 
So, such a nonperturbative flavor breaking is expected to be reflected also in the topological susceptibility.

Actually, in Ref.~\cite{Kawaguchi:2020kdl} the authors 
demonstrated a significant flavor breaking effect on 
the topological susceptibility at high temperatures 
around and above the pseudo-critical temperature, which 
was shown to be generated from nonperturbative thermal 
loop corrections to quark condensates, that are flavorful 
for light and strange quarks. 
There it was emphasized that the presence of 
strange quark contribution 
would be nonperturbatively crucial, which cannot be 
realized by the conventional chiral perturbation theory (ChPT), implying that it drives 
the $U(1)_A$ anomaly effect to be sizable even at high temperatures.

In this paper, we develop the analysis done in~\cite{Kawaguchi:2020kdl}, 
by further investigating the flavor 
breaking effects on meson spectroscopy around 
the hot QCD criticality. 
We also supply the details to reach the result reported in the literature.

As was done in~\cite{Kawaguchi:2020kdl}, 
we work on a three-flavor linear-sigma model 
including an $SU(3)$ flavor violation induced by $U(1)_A$ anomaly which we call an axial-anomaly induced-flavor breaking term~\cite{Kuroda:2019jzm}, 
and employ a nonperturbative analysis following 
the Cornwall-Jackiw-Tomboulis (CJT) formalism~\cite{Cornwall:1974vz}. 
The nonperturbative flavor breaking is provided 
from two ingredients: one is nonperturbative thermal 
corrections, and the other is an axial anomaly-induced 
flavor breaking interaction.

The model is shown to undergo a chiral crossover 
with a pseudo-critical temperature, consistently with lattice observations.  
Nonpertubative computation of thermal loop corrections shows that up-and down-quark condensates drop 
faster than the strange quark's toward the criticality, but still keep nonzero value 
even going far above the critical temperature, 
which implies the nonperturbative generation of a significant flavor breaking. It turns out that the introduced anomaly-related flavor-breaking effect acts as a catalyzer for the chiral restoration, while it reduces 
the amount of flavor breaking in the up, down and strange condensates.

Nonperturbative thermal corrections also generate 
a dramatic deformation for the meson flavor mixing structures, in passing the chiral 
crossover, in which the anomaly-induced favor breaking is found to be almost irrelevant. 
Of interest is that the meson spectroscopy gets corrected by the net nonperturbative flavor breaking effects around and above the critical temperature, where the scalar meson mass hierarchy is significantly altered by the presence of the anomaly-related flavor breaking.

We also pay particular attention to the topological susceptibility, with the correct identification keeping 
the flavor universal nature for the current quark masses.
We find that around and above 
the chiral crossover, the topological susceptibility gets 
a nonperturbative flavor breaking via the surviving strange quark condensate, 
which cannot be detected by the chiral perturbation 
theory (ChPT), and significantly deviates from the dilute instanton gas prediction. 
There 
the anomaly-induced flavor breaking plays a role 
of the destructive interference for the enhancement, 
as in the flavor breaking in the quark condensates. 

Remarkably,  
the $U(1)_A$ breaking dictated by nonzero topological susceptibility is catalyzed 
by the nonperturbative strange quark condensate 
at around the chiral crossover criticality, which may account for the tension in the effective restoration of the $U(1)_A$ symmetry
currently observed on lattices near the chiral limit 
with two flavors~\cite{Cossu:2013uua,Tomiya:2016jwr,Suzuki:2019vzy,Suzuki:2020rla}
and 2+1 flavors~\cite{Bazavov:2012qja,Buchoff:2013nra,Ding:2019prx}.


\section{A linear sigma model at zero temperature}
In this section, we begin by giving a brief review of the three-flavor linear-sigma model involving an axial-anomaly induced-flavor breaking proposed in the literature~\cite{Kuroda:2019jzm}. 
This model with the introduced flavor-breaking axial anomaly 
has been shown to successfully 
reproduce what is called the 
inverse mass hierarchy for scalar mesons lighter than 1 GeV ($m_{f_0(500)} < m_{\kappa(700)=K_0^*(700)} <m_{a_0(980)} (\simeq m_{f_0(980)})$).  
The model is built based on the chiral symmetry -- 
which is spontaneously 
broken by vacuum expectation values of the sigma model field in 
a way consistent with the underlying QCD -- with including 
the minimal flavor violation effect induced from the 
current quark mass difference. 

\subsection {Three-flavor linear-sigma model involving the axial-anomaly induced-flavor breaking}\label{linear_model_k}

We start with introducing the building block, 
the linear sigma model field $\Phi$ as a  $3\times 3$ matrix field $\Phi$. 
It is parametrized by the scalar- and pseudoscalar-meson nonets as 
\begin{eqnarray}
\Phi&=&\Phi_aT_a=(\sigma_a+i\pi_a)T_a,
\label{PhiField}
\end{eqnarray}
where $\sigma_a$ are the scalar fields and $\pi_a$ are the pseudoscalar fields. $T_a=\lambda_a/2$ ($a=0,1,\cdots,8$) are the generators of $U(3)$, where $\lambda_{a=1,\cdots,8}$ are the Gell-Mann matrices with $\lambda_0=\sqrt{2/3}\,{\bm 1}_{3\times 3}$.
The generators satisfy the following identities,
\begin{eqnarray}
{\rm tr}[T^aT^b]&=&\frac{1}{2}\delta^{ab},\nonumber\\
{[}T^a,T^b]&=&if_{abc} T_c,\nonumber\\
\{T_a,T_b\}&=&d_{abc}T_c,
\end{eqnarray}
where 
$f_{abc}$ and $d_{abc}$ (for $a,b,c=1,\cdots,8$) are the antisymmetric and symmetric structure constants of $SU(3)$ respectively, 
and the following conditions are satisfied 
\begin{eqnarray}
f_{ab0}=0,\;\;\;
d_{ab0}=\sqrt{\frac{2}{3}}\delta_{ab}.
\end{eqnarray}
Under the chiral $SU(3)_L\times SU(3)_R\times U(1)_A$ symmetry,
the $3\times 3$ matrix $\Phi$ transforms as 
\begin{eqnarray}
\Phi\to g_A\cdot g_L\cdot \Phi\cdot g_R^\dagger.
\end{eqnarray}
where 
$g_{L,R}\in SU(3)_{L,R}$ and $g_A\in U(1)_A$.

The scalar mesons ($J^P=0^+$) and the pseudoscalar mesons ($J^P=0^-$) are embedded 
in the $\sigma_a T_a$ and $\pi_a T_a$, respectively, like 
\begin{eqnarray}
\sigma_a T_a&=&
\frac{1}{\sqrt{2}}
\begin{pmatrix}
\frac{a_0}{\sqrt{2}}+\frac{\sigma_8}{\sqrt{6}}+\frac{\sigma_0}{\sqrt{3}}&a^+&\kappa^+\\
a^-&-\frac{a_0}{\sqrt{2}}+\frac{\sigma_8}{\sqrt{6}}+\frac{\sigma_0}{\sqrt{3}} &\kappa^0\\
\kappa^-&\bar \kappa^0&-\frac{2\sigma_8}{\sqrt{6}}+\frac{\sigma_0}{\sqrt{3}}
\end{pmatrix},\nonumber\\
\pi_a T_a&=&
\frac{1}{\sqrt{2}}
\begin{pmatrix}
\frac{\pi^0}{\sqrt{2}}+\frac{\eta_8}{\sqrt{6}}+\frac{\eta_0}{\sqrt{3}}&\pi^+&K^+\\
\pi^-&-\frac{\pi^0}{\sqrt{2}}+\frac{\eta_8}{\sqrt{6}}+\frac{\eta_0}{\sqrt{3}} &K^0\\
K^-&\bar K^0&-\frac{2\eta_8}{\sqrt{6}}+\frac{\eta_0}{\sqrt{3}}
\end{pmatrix},
\end{eqnarray}
where
$\pi^\pm=(\pi_1\mp i\pi_2)/\sqrt{2}$ and $\pi^0=\pi_3$ are pions,
$K^\pm=(\pi_4\mp i\pi_5)/\sqrt{2}$, $K^0=(\pi_6- i\pi_7)/\sqrt{2}$ and
$\bar K^0=(\pi_6+ i\pi_7)/\sqrt{2}$ are kaons,
$\eta_0=\pi_0$ and $\eta_8=\pi_8$ are admixtures of the $\eta$ meson and $\eta'$.
In the scalar meson nonet, the $a_0(980)$ meson is identified as 
the isotriplet component, $a^\pm=(\sigma_1\mp i\sigma_2)/\sqrt{2}$ and $a_0=\sigma_3$. 
The $\kappa(700)=K_0^*(700)$ mesons, forming the isodoublet (including single strange quark as valence quarks), 
are assigned as 
$\kappa^\pm=(\sigma_4\mp i\sigma_5)/\sqrt{2}$, $\kappa^0=(\sigma_6- i\sigma_7)/\sqrt{2}$ and
$\bar \kappa^0=(\sigma_6+ i\sigma_7)/\sqrt{2}$. 
The $\sigma_0$ and $\sigma_8$ are referred to as the admixtures of the $f_0(500)$ and $f_0(980)$.

With the linear sigma model field $\Phi$ as the 
building block at hand,
the three-flavor linear sigma model 
is written down
\begin{eqnarray}
{\cal L}={\rm tr}\left[ \partial_\mu \Phi\partial^\mu \Phi^\dagger \right]-V.
\label{linear_sigma_model}
\end{eqnarray}
where $V$ represents the potential terms,
\begin{eqnarray}
V&=&V_0+V_{\rm anom}+V_{\rm anom}+V_{\rm SB}+V_{\rm SB-anom}.
\label{pot}
\end{eqnarray}

$V_0$ is an invariant part under the $SU(3)_L\times SU(3)_R\times U(1)_A$ symmetry,
in which the mass term of $\Phi$ and the four-point interaction terms are incorporated,
\begin{eqnarray}
V_0=\mu^2{\rm tr }[(\Phi^\dagger\Phi)]
+\lambda_1{\rm tr }[(\Phi^\dagger\Phi)^2]
+\lambda_2({\rm tr }[(\Phi^\dagger\Phi)])^2,
\label{mu_lam12}
\end{eqnarray}
where $\mu^2$ can take either a positive or negative value and $\lambda_{1,2}$ are dimensionless quartic coupling constants.

The $U(1)_A$ anomalous part, but keeping 
the chiral $SU(3)_L \times SU(3)_R$ symmetry,  
is incorporated in $V_{\rm anom}$.
The lowest dimensional operator for the $U(1)_A$ anomalous term is given by a la Kobayashi-Maskawa-'t Hooft (KMT) \cite{Kobayashi:1970ji,Kobayashi:1971qz,tHooft:1976rip,tHooft:1976snw}, 
\begin{eqnarray}
V_{\rm anom}=-B\left({\rm det}[\Phi]+{\rm det}[\Phi^\dagger]\right),
\label{detterm}
\end{eqnarray}
where the parameter $B$ is real, having mass dimension one. 
This $V_{\rm anom}$ is necessary to supply 
the sufficiently large mass for the $\eta'$ meson to be no longer a Nambu-Goldstone boson.

The explicit chiral symmetry breaking originates from the current quark mass matrix ${\cal M}={\rm diag}( m_u,  m_d, m_s)$ in the underlying QCD Lagrangian. 
We incorporate this explicit breaking structure into the linear sigma model, by regarding the current quark mass matrix  ${\cal M}$ as 
a spurion field, which 
transforms in the same way as $\Phi$ does, ${\cal M}\to g_A\cdot g_L\cdot{\cal M}\cdot g_R^\dagger$ and 
develops the vacuum expectation value 
$\langle {\cal M} \rangle={\rm diag}\{ m_u,  m_d, m_s\}$.  
Thus the explicit breaking terms can be introduced in 
the linear-sigma model-Lagrangian in a chiral-invariant way. 
In the minimal flavor violation limit where only the one 
${\cal M}$ can be operative, the $V_{\rm SB}$ part, 
free from the nonperturbative axial anomaly, thus goes like  
\begin{eqnarray}
V_{\rm SB}=-c{\rm tr}[{\cal M}\Phi^\dagger+{\cal M}^\dagger \Phi],
\end{eqnarray}
where the parameter $c$ is taken to be real, and has mass dimension two. 
The parameter $c$ will not solely show up in the physical quantities, because its degree of freedom corresponds to the renormalization scale ambiguity in defining quark condensates. 

The $V_{\rm SB-anom}$ is the axial-anomaly induced-flavor breaking term, 
introduced in~\cite{Kuroda:2019jzm}.
In the minimal flavor violation limit where the single ${\cal M}$ is only allowed to be inserted, the $V_{\rm SB-anom}$ is cast is into the form~\footnote{
A field redefinition for the linear sigma model field $\Phi$ 
(or the spurion field ${\cal M}$)  
can eliminate the $k$-term, however,
instead would necessarily yield terms including 
higher orders in ${\cal M}$. 
Hence, the field redefinition will be off from the criterion of 
the minimal flavor violation, on which the present model is based. 
On this basis, it is robust that the $k$ term is only the source related 
to the anomaly-related flavor breaking. 
Though being not associated with the U(1) axial anomaly
(which turns out to be manifest in the limit of nonlinear realization, i.e.,when reduced back to the chiral perturbation theory), 
introduction of terms like ${\rm tr}[\mathcal{M}\Phi^\dagger M \Phi^\dagger + h.c.]$ and
${\rm tr}[\mathcal{M}\Phi^\dagger]{\rm tr}[\Phi\Phi^\dagger] + h.c.$, will not 
be contradicted with our criterion of the minimal flavor violation, 
hence might make some effects on what we have argued on the inverse mass hierarchy, in addition to the $k$-term. 
The extended analysis including those operators will be 
pursued elsewhere. 
}, 
\begin{eqnarray}
V_{\rm SB-anom}=-kc\left[\epsilon_{abc}\epsilon^{def}{\cal M}^a_d\Phi^b_e \Phi^c_f+{\rm h.c.}
\right],
\end{eqnarray}
where the parameter $k$ is real, having mass dimension -1,  and $\epsilon_{abc}$ is totally antisymmetric tensor
under the exchange of indices $a, b$ and $c$ with $\epsilon_{123}=1$. 
It has been demonstrated~\cite{Kuroda:2019jzm} 
that, because of the flavor-singlet feature, 
this $k$ term 
supplies the contribution $\propto m_s$ only to 
the mass of $a_0(980)$ meson, while the $K_0^*(700)$ gets 
the term $\propto m_{u,d}$, 
so 
it plays the crucial role of realizing 
the inverse mass hierarchy for 
the scalar mesons lighter than $1$ GeV:  
$m[a_0(980)]\simeq m[f_0(980)]>m[K_0^*(700)]>m[f_0(500)]$. 

\subsection{Spontaneous chiral symmetry breaking, 
meson masses, and topological susceptibility at zero temperature}\label{spo_mass_T0}

In the linear sigma model,
the spontaneous chiral symmetry breaking occurs by 
the nonzero vacuum expectation value of the $\Phi$ field: 
\begin{eqnarray}
\langle \Phi \rangle=\bar \Phi=T_a\bar \sigma_a.
\end{eqnarray}
We impose the isospin symmetry, $m_l \equiv 
m_u=m_d\neq m_s$, 
so that the vacuum expectation values $\bar{\sigma}_a$ 
are taken as
\begin{eqnarray}
\bar\sigma_a T_a&=&\bar\sigma_0 T_0+\bar\sigma_8 T_8\nonumber\\
&=&
{\rm diag}(\bar \Phi_1,\bar\Phi_1,\bar\Phi_3)
\label{vevsigma}
\end{eqnarray}
where
\begin{eqnarray}
\bar\Phi_1&=&
\frac{1}{\sqrt{6}}\bar\sigma_0+\frac{1}{2\sqrt{3}}\bar\sigma_8,
\nonumber\\
\bar\Phi_3&=&
\frac{1}{\sqrt{6}}\bar\sigma_0-\frac{1}{\sqrt{3}}\bar\sigma_8.
\end{eqnarray} 
For the background fields $\bar\sigma_{0,8}$,  
the potential Eq.~(\ref{pot}) 
is evaluated as  
\begin{eqnarray}
V(\bar \sigma)&=&\frac{1}{2}\mu^2(\bar \sigma_0^2+\bar \sigma_8^2)
+\lambda_1\left[
\frac{1}{12}\bar \sigma_0^4+\frac{1}{2}\bar \sigma_0^2\bar \sigma_8^2-\frac{1}{3\sqrt{2}}\bar \sigma_0\bar \sigma_8^3
+\frac{1}{8}\bar \sigma_8^4
\right]
+\frac{\lambda_2}{4}(\bar \sigma_0^2+\bar \sigma_8^2)^2\nonumber\\
&&-B\left[
\frac{1}{3\sqrt{6}}\bar \sigma_0^3-\frac{1}{2\sqrt{6}}\bar \sigma_0\bar \sigma_8^2
-\frac{1}{6\sqrt{3}}\bar \sigma_8^3
\right]
-\sqrt{\frac{2}{3}}(2cm_l+cm_s)\sigma_0
-\frac{2}{\sqrt{3}}(cm_l-cm_s)\sigma_8
\nonumber\\
&&
-kc\left[
\frac{m_l}{3}(4\bar\sigma_0^2-2{\sqrt{2}}\bar\sigma_0\bar\sigma_8-4\bar\sigma_8^2  )
+\frac{m_s}{3}(2\bar\sigma_0^2+2\sqrt{2}\bar\sigma_0\bar\sigma_8+\bar\sigma_8^2  )
\right].
\label{TreePot}
\end{eqnarray}
From this potential, the stationary conditions for $\bar \sigma_0$ and $\bar \sigma_8$ read  
\begin{eqnarray}
c\sqrt{\frac{2}{3}}(2m_l+m_s)
&=&
\bar\sigma_0\left[
\mu^2+\frac{\lambda_1}{3}(\bar\sigma_0)^2+\lambda_2(\bar\sigma_0)^2-\frac{B}{\sqrt{6}}\bar\sigma_0
\right]
+(\bar\sigma_8)^2\left[
\lambda_1\bar\sigma_0+\lambda_2\bar\sigma_0-\frac{\lambda_1}{3\sqrt{2}}\bar\sigma_8+\frac{B}{2\sqrt{6}}
\right]\nonumber\\
&&
-kc\left[
\frac{m_l}{3}(8\bar\sigma_0-2{\sqrt{2}}\bar\sigma_8)
+\frac{m_s}{3}(4\bar\sigma_0+2\sqrt{2}\bar\sigma_8)
\right],\nonumber\\
c\frac{2}{\sqrt{3}}(m_l-m_s)&=&
\bar\sigma_8\left[\mu^2+\frac{B}{\sqrt{6}}\bar\sigma_0
+\frac{B}{2\sqrt{3}}\bar\sigma_8+(\lambda_1+\lambda_2)\bar\sigma_0^2
-\frac{\lambda_1}{\sqrt 2}\bar\sigma_0\bar\sigma_8+
\left(\frac{\lambda_1}{2}+\lambda_2\right)(\bar\sigma_8)^2
\right]\nonumber\\
&&
-kc\left[
\frac{m_l}{3}(-2{\sqrt{2}}\bar\sigma_0-8\bar\sigma_8  )
+\frac{m_s}{3}(2\sqrt{2}\bar\sigma_0+2\bar\sigma_8  )
\right].
\label{tree_sta_con}
\end{eqnarray}

\subsubsection{Scalar and pseudoscalar meson masses}

Around the vacuum given by $\bar\sigma_{0,8}$ in Eq.(\ref{tree_sta_con}),  
the scalar mesons arise as the fluctuating modes, to form 
the mass matrix $(m_S^2)_{ij}$ ($i,j=0,1,4,8$), with the matrix elements,  
\begin{eqnarray}
(m_S^2)_{00}&=&
\mu^2+\lambda_1(\bar\sigma_0^2+\bar\sigma_8^2)
+\lambda_2(3\bar\sigma_0^2+\bar\sigma_8^2)
-B\sqrt{\frac{2}{3}}\bar\sigma_0
-\frac{4}{3}kc(2m_l+m_s),\nonumber\\
(m_S^2)_{88}&=&
\mu^2+
\lambda_1\left[\bar\sigma_0^2-\sqrt{2}\bar\sigma_0\bar\sigma_8
+\frac{3}{2}\bar\sigma_8^2
\right]+
\lambda_2(\bar\sigma_0^2+3\bar\sigma_8^2)
+\frac{B}{\sqrt{3}}\left(\frac{1}{\sqrt2}\bar\sigma_0+\bar\sigma_8\right)
+\frac{2}{3}kc(4 m_l-m_s),
\nonumber\\
(m_S^2)_{08}&=&
\lambda_1\left[2\bar\sigma_0\bar\sigma_8
-\frac{1}{\sqrt{2}}\bar\sigma_8^2\right]
+2\lambda_2\bar\sigma_0\bar\sigma_8
+\frac{B}{\sqrt{6}}\bar\sigma_8
+\frac{2\sqrt{2}}{3}kc( m_l-m_s),
\label{TreeMassS} \\ 
(m_S^2)_{11}&=&
\mu^2+
\lambda_1\left[\bar\sigma_0^2+\sqrt{2}\bar\sigma_0\bar\sigma_8
+\frac{1}{2}\bar\sigma_8^2\right]+
\lambda_2(\bar\sigma_0^2+\bar\sigma_8^2)
+\frac{B}{\sqrt{3}}\left(\frac{1}{\sqrt2}\bar\sigma_0-\bar\sigma_8\right)
+2kcm_s,
\nonumber\\
(m_S^2)_{44}&=&\mu^2+
\lambda_1\left[\bar\sigma_0^2-\frac{1}{\sqrt{2}}\bar\sigma_0\bar\sigma_8
+\frac{1}{2}\bar\sigma_8^2\right]+
\lambda_2(\bar\sigma_0^2+\bar\sigma_8^2)
+\frac{B}{\sqrt{3}}\left(\frac{1}{\sqrt2}\bar\sigma_0+\frac{1}{2}\bar\sigma_8\right)
+2kc m_l,\nonumber
\end{eqnarray}
where $(m_S^2)_{11}$ and $(m_S^2)_{44}$ are identified as the $a_0(980)$ meson and the $K_0^*(700)$ respectively,
$m_{a_0}^2\equiv(m_S^2)_{11}$, $m_{\kappa}^2\equiv (m_S^2)_{44}$.
As shown in Eq.~(\ref{TreeMassS}),  the $(m_S^2)_{11}$ has the $2kcm_s$ term, which comes from the $V_{\rm SB-anom}$ term. 
On the other hand, the contribution to $(m_S^2)_{44}$ from the  $V_{\rm SB-anom}$ term is given as the $2kcm_l$ term. 
Due to $m_s>m_l$ 
the axial-anomaly induced-flavor breaking $k$-term supplies the sufficient contribution to  the $a_0(980)$ meson,
so that the $a_0$ meson is necessarily heavier 
than the $K_0^*(700)$ meson, 
hence the desired inverse mass hierarchy $m_{a_0}>m_{\kappa}$ 
is realized~\cite{Kuroda:2019jzm}.

The mixing part 
for $(m_S^2)_{00}$ and $(m_S^2)_{88}$ 
can be diagonalized by an orthogonal transformation in the following way:
\begin{eqnarray}
\tilde \sigma_i&=&\left(O^{-1}_S\right)_{ia}\sigma_a,\nonumber\\
{[}\tilde m_{S}^2]_{(i)}\delta_{ij}&=& \left(O^{-1}_{S}\right)_{ia} [m_{S}^2]_{ab}\left(O_{S}\right)_{bj},
\label{transformT0S}
\end{eqnarray}
where ${[}\tilde m_{S}^2]_{(i)}$ are the 
mass eigenvalues.  
Thus the $f_0(500)$ and $f_0(980)$ masses are assigned and evaluated as 
\begin{eqnarray}
m^2[f_0(500)]&=&(\tilde m_S ^2)_{(0)}=(m_S^2)_{00}\cos^2\theta_S^0+(m_S^2)_{88}\sin^2\theta_S^0
-2(m_S^2)_{08}\cos\theta_S^0 \sin\theta_S^0,
\nonumber\\
m^2[f_0(980)]&=&(\tilde m_S ^2)_{(8)}=(m_S^2)_{00}\sin^2\theta_S^0+(m_S^2)_{88}\cos^2\theta_S^0
+2(m_S^2)_{08}\cos\theta_S^0 \sin\theta_S^0,
\end{eqnarray}
where $\theta^0_S$ represents the scalar mixing angle 
\begin{eqnarray}
\theta_{S}^0=\frac{1}{2}\arctan\left[
\frac{2(m_{S}^2)_{08}}{(m_{S}^2)_{00}-(m_{S}^2)_{88}}\right].
\end{eqnarray}

In a way similar to the scalar meson sector, 
the pseudoscalar mesons are also in part mixed 
to form the mass matrix $(m_{P}^2)_{ij}$ ($i,j,=0,1,4,8$) with 
the matrix elements, 
\begin{eqnarray} 
(m_P^2)_{00}&=&
\mu^2+
\frac{\lambda_1}{3}(\bar\sigma_0^2+\bar\sigma_8^2)
+\lambda_2(\bar\sigma_0^2+\bar\sigma_8^2)
+B\sqrt{\frac{2}{3}}\bar\sigma_0
+\frac{4}{3}kc(2 m_l+m_s),
\nonumber\\
(m_P^2)_{11}&=&
\mu^2+
\lambda_1\left[\frac{1}{3}\bar\sigma_0^2+\frac{\sqrt{2}}{3}\bar\sigma_0\bar\sigma_8
+\frac{1}{6}\bar\sigma_8^2\right]+
\lambda_2(\bar\sigma_0^2+\bar\sigma_8^2)
-\frac{B}{\sqrt{3}}\left(\frac{1}{\sqrt2}\bar\sigma_0-\bar\sigma_8\right)
-2kcm_s,\nonumber\\
(m_P^2)_{44}&=&
\mu^2+
\lambda_1\left[\frac{1}{3}(\bar\sigma_0)^2-\frac{1}{3\sqrt{2}}\bar\sigma_0\bar\sigma_8
+\frac{7}{6}(\bar\sigma_8)^2\right]+
\lambda_2(\bar\sigma_0^2+\bar\sigma_8^2)
-\frac{B}{\sqrt{3}}\left(\frac{1}{\sqrt2}\bar\sigma_0+\frac{1}{2}\bar\sigma_8\right)
-2kc m_l,
\nonumber\\
(m_P^2)_{88}&=&
\mu^2+
\lambda_1\left[\frac{1}{3}(\bar\sigma_0)^2-\frac{\sqrt{2}}{3}\bar\sigma_0\bar\sigma_8
+\frac{1}{2}(\bar\sigma_8)^2
\right]+
\lambda_2(\bar\sigma_0^2+\bar\sigma_8^2)
-\frac{B}{\sqrt{3}}\left(\frac{1}{\sqrt2}\bar\sigma_0+\bar\sigma_8\right)
-\frac{2}{3}kc(4m_l-m_s),
\nonumber\\
(m_P^2)_{08}&=&
\frac{2}{3}\lambda_1\bar\sigma_0\bar\sigma_8
-\frac{\lambda_1}{3\sqrt{2}}\bar\sigma_8^2
-\frac{B}{\sqrt{6}}\bar\sigma_8
-\frac{2\sqrt{2}}{3}kc( m_l-m_s),
\label{TreeMassP}
\end{eqnarray}
where $(m_P^2)_{11}$ and $(m_P^2)_{44}$ are identified as the pion mass and the kaon mass, respectively, 
$m_{\pi}^2\equiv(m_P^2)_{11}$, $m_{K}^2\equiv (m_P^2)_{44}$. 
As the consequence of the chiral partner structure in the linear sigma model,  
The $\pi_0$ and $\pi_8$ mix in a similar way to the scalar meson's.  The mass matrix can be diagonalized by the orthogonal transformation,
\begin{eqnarray}
\tilde \pi_i&=&\left(O^{-1}_P\right)_{ia}\pi_a,\nonumber\\
{[}\tilde m_{P}^2]_{(i)}\delta_{ij}&=& \left(O^{-1}_{P}\right)_{ia} [m_{S,P}^2]_{ab}\left(O_{P}\right)_{bj},
\label{transformT0P}
\end{eqnarray}
where the ${[}\tilde m_{P}^2]_{(i)}$ are the mass eigenvalues. 
Thus, the $\eta'$ and $\eta$ meson masses are assigned and evaluated  as
\begin{eqnarray} 
m^2_{\eta'}&=&(\tilde m_P ^2)_{(0)}=(m_P^2)_{00}\cos^2\theta_P^0+(m_P^2)_{88}\sin^2\theta_P^0
-2(m_P^2)_{08}\cos\theta_P^0 \sin\theta_P^0,
\nonumber\\
m^2_\eta&=&(\tilde m_P ^2)_{(8)}=(m_P^2)_{00}\cos^2\theta_P^0+(m_P^2)_{88}\sin^2\theta_P^0
+2(m_P^2)_{08}\cos\theta_P^0 \sin\theta_P^0,
\end{eqnarray}
where $\theta_P^0$ denotes the pseudoscalar mixing angle at the zero temperature, which is given by 
\begin{eqnarray}
\theta_{P}^0=\frac{1}{2}\arctan\left[
\frac{2(m_{P}^2)_{08}}{(m_{P}^2)_{00}-(m_{P}^2)_{88}}
\right].
\end{eqnarray}




By using the stationary condition in Eq.~(\ref{tree_sta_con}),
the pion mass and the kaon mass can be written as
\begin{eqnarray}
m_\pi^2&=&\frac{cm_l}{\bar\Phi_1}(1+2k\bar \Phi_3),\nonumber\\
m_K^2&=&\frac{cm_l+cm_s}{\bar \Phi_1+\bar \Phi_3}(1+2k\bar\Phi_1).
\label{pionkaonmass}
\end{eqnarray} 
It looks like that those masses significantly get corrections  
from the $k$ term having an intrinsic-flavor violation-nature 
associated with the determinant form (breaking the intrinsic parity).  So, one might doubt if the present model surely 
satisfies the low-energy theorem for three-flavor case, 
such as the  Gell-Mann-Oakes-Renner relations. 
To check this point, we shall look at  
the quark condensates, and 
the pion decay constant $f_\pi$ and the kaon decay constant $f_K$. 
The former reads  
\begin{eqnarray}
\langle \bar ll \rangle &=&\frac{\partial V(\bar \sigma) }{\partial m_l}=-2c(\bar \Phi_1+2k\bar \Phi _1\bar \Phi_3),\nonumber\\
\langle \bar ss \rangle&=&\frac{\partial V(\bar \sigma) }{\partial m_s}=-2c(\bar \Phi_3+2k\bar \Phi_1^2), 
\label{condensates_tree}
\end{eqnarray} 
while the latter are read off from the overlap amplitudes between  
the corresponding axialvector currents and pseudoscalars,  
to be given as 
\begin{eqnarray}
f_\pi&=&2\bar\Phi_1,\nonumber\\
f_K&=&\bar\Phi_1+\bar\Phi_3.
\label{decay_constants}
\end{eqnarray} 
From these, we find 
\begin{eqnarray}
f_\pi^2 m_\pi^2&=&(m_l+m_d)\langle \bar ll \rangle,\nonumber\\
f_K^2 m_K^2&=&\frac{m_l+m_d}{2}(\langle \bar ll \rangle+\langle \bar ss \rangle),
\label{GMOR}
\end{eqnarray} 
which are precisely the Gell-Mann-Oakes-Renner relations. 
Note an accidental cancellation of the $k$ dependence 
arising from both quark condensates and pion, kaon masses.  
Thus, the low energy theorem is intact  
even in the presence of the intrinsically flavor-violating $k$-term~\cite{Kuroda:2019jzm}.

\subsubsection{Topological susceptibility: flavor-universal nature}

We next turn to topological susceptibility $\chi_{\rm top}$. 
In hot QCD, the $\chi_{\rm top}$ is a crucial quantity to measure the topological charge fluctuation of the QCD $\theta$-vacuum. 
It is defined as the curvature of the free energy of QCD with respect to the $\theta$ at the QCD vacuum with $\theta=0$: 
\begin{eqnarray}
\chi_{\rm top}=
-\int_T d^4x \frac{\delta^2 V(\theta)}{\delta \theta(x)\delta \theta(0)}\Biggl |_{\theta=0}.
\label{topsus_dif}
\end{eqnarray}
where $V(\theta)$  is the $\theta$-dependent vacuum energy
\textcolor{black}{and the temperature integral is denoted as
$\int_T d^4x\equiv \int^{1/T}_0d\tau d^3\vec x$}
.
Furthermore, the QCD $\theta$-vacuum is correlated with the 
presence of the $U(1)_A$ anomaly including quark mass contributions. 
As will be seen later, the $\chi_{\rm top}$ must have 
a flavor-universal nature, and vanish when either of all current quark masses goes to zero, because the nonperturbative $U(1)_A$ anomaly induced 
by topological gluon configurations can always be transferred to current quark mass terms in the QCD generating functional, once the 
current quark masses are introduced in the theory. 
Thereby, the $\chi_{\rm top}$ also plays an important role of monitoring some (effective or partial) $U(1)_A$ restoration in the presence of nonzero current quark masses. (That might or might not happen at high temperatures, 
as will be seen later.)   
In this section, we pay our attention to this topological susceptibility in the present linear sigma model with the anomaly-induced flavor breaking, the $k$-term.

Generically, 
the $\theta$-parameter can be introduced 
through matching the QCD generating functional with the one corresponding  to the low-energy effective model which one works on. 
So, we shall start from the Euclidean generating functional of QCD 
given as a function of $\theta$, 
\begin{eqnarray}
Z_{\rm QCD}&=&\int [\Pi_f dq_f d\bar q_f] [dA]
\exp\Biggl[-\int_T d^4x \Biggl\{
\sum_f\left(
\bar q^f_Li\gamma^\mu D_\mu q^f_L
+
\bar q^f_Ri\gamma^\mu D_\mu q^f_R
+\bar q^f_L m_f q^f_R+\bar q^f_R m_f q^f_L\right)
\nonumber\\
&&
+\frac{1}{4g^2}(F_{\mu\nu}^a)^2 
+\frac{i\theta}{32\pi^2}F_{\mu\nu}^a\tilde F_{\mu\nu}^a
\Biggl\}
\Biggl],
\label{QCDgene}
\end{eqnarray}
where
$q^f_{L(R)}$ denote the left- (right-) handed quark fields; the covariant derivative of the quark field is represented as $D_\mu$ involving the gluon fields $A$; 
$F_{\mu\nu}^a$ is the field strength of the gluon fields with 
$g$ being the QCD coupling constant. 
The finiteness of the $\theta$-parameter indicates that the QCD system is put in the CP violation domain.

Under the $U_A(1)$ rotation with the rotation angle $\theta_f$, 
the left- and right-handed quark fields are transformed as 
\begin{eqnarray} 
q^f_L&\to& \exp\left( -i\theta_f/2 \right)q^f_L, \nonumber\\
q^f_R&\to& \exp\left( i\theta_f/2 \right)q^f_R, 
\end{eqnarray}
By rotating the chiral quark fields as above, one finds that 
the extra phase factor shows up in the QCD generating functional:  
\begin{eqnarray}
&&\int [\Pi_f dq_f d\bar q_f] [dA]
\exp\Biggl[-\int_T d^4x \Biggl\{
\sum_f\left(
\bar q^f_Li\gamma^\mu D_\mu q^f_L
+
\bar q^f_Ri\gamma^\mu D_\mu q^f_R
+
\bar q^f_L m_f e^{i\theta_f} q^f_R+\bar q^f_R m_f e^{-i\theta_f} q^f_L\right)
\nonumber\\
&&
+\frac{1}{4g^2}(F_{\mu\nu}^a)^2
+
\frac{i(\theta-\bar\theta)}{32\pi^2}
F_{\mu\nu}^a\tilde F_{\mu\nu}^a
\Biggl\}
\Biggl],
\end{eqnarray}
where 
$\bar\theta=\sum_{f=u,d,s} \theta_f = \theta_u+\theta_d+\theta_s$.
Thus, in the presence of nonzero current quark masses, 
generically the $\theta$-term ($\theta F_{\mu\nu}^a\tilde F_{\mu\nu}^a$) can be rotated away from the QCD generating functional by choosing the chiral rotation angle like, 
\begin{eqnarray}
\theta=\bar \theta=\theta_u+\theta_d+\theta_s,
\label{choicetheta_f}
\end{eqnarray}
Instead, the $\theta$-dependence is fully absorbed into the quark mass matrix. 
Crucial is then to note that 
if either of quarks are massless, 
 the $\theta$-dependence can be completely rotated away from the QCD generating functional. 
It implies that for small enough $\theta_f$, 
the $\theta_f$ are constrained by the flavor singlet condition~\cite{Baluni:1978rf},
\begin{eqnarray}
m_u\theta_u=m_d\theta_d=m_s\theta_s \equiv x.
\label{singlet_con}
\end{eqnarray}
By using this flavor singlet condition 
with the choice of the parameter $\bar \theta$ in Eq.~(\ref{choicetheta_f}),
the parameter $\theta_f$s are determined as
\begin{eqnarray}
\theta_u=\frac{\bar m}{m_u} \theta,\;\;\;
\theta_d=\frac{\bar m}{m_d} \theta,\;\;\;
\theta_s=\frac{\bar m}{m_s} \theta,
\label{conspara}
\end{eqnarray}
where
\begin{eqnarray}
\bar m=\left(\frac{1}{m_u}+\frac{1}{m_d}+\frac{1}{m_s} \right)^{-1}.
\end{eqnarray}

Thus the $\theta$-dependent vacuum energy of QCD 
is cast into the form 
\begin{eqnarray}
V_{\rm QCD}(\theta)&=&
-\ln\left[
\int [\Pi_f dq_f d\bar q_f] [dA]
\exp\left(-\int_T d^4x 
{\cal L}_{\rm QCD}^{(\theta)}
\right)\right]
\end{eqnarray}
where
\begin{eqnarray}
{\cal L}_{\rm QCD }^{(\theta)}
=
\sum_f\left(
\bar q^f_Li\gamma^\mu D_\mu q^f_L
+
\bar q^f_Ri\gamma^\mu D_\mu q^f_R
\right)
+
\bar q_L {\cal M}_\theta q_R+\bar q_R {\cal M}_\theta^\dagger q_L
+\frac{1}{4g^2}(F_{\mu\nu}^a)^2
\label{TQCDlag}
\end{eqnarray}
with 
${\cal M}_\theta$ being the $\theta$-dependent quark matrix,
\begin{eqnarray}
{\cal M}_\theta=
{\rm diag}\left[m_u\exp\left(i\frac{\bar m}{m_u}\theta\right), m_d\exp\left(i\frac{\bar m}{m_d}\theta\right), m_s\exp\left(i\frac{\bar m}{m_s}\theta\right)\right].
\label{calM-theta}
\end{eqnarray}
Therefore, 
from Eq.~(\ref{topsus_dif})
the topological susceptibility of QCD is evaluated as
\begin{eqnarray}
\chi_{\rm top}^{({\rm QCD})}
&=&
-\int_T d^4x \frac{\delta^2 V_{\rm QCD}}{\delta \theta(x)\delta \theta(0)}\Biggl |_{\theta=0}
\nonumber\\
&=&
\left(\frac{\langle \bar uu \rangle(T)}{m_u}+\frac{\langle \bar dd \rangle(T)}{m_d}+\frac{\langle \bar ss \rangle(T)}{m_s}
\right)\bar m^2
+
\int_T d^4x\langle\Bigl(\sum_fi\bar q^f(x)\gamma_5 q^f(x)\Bigl)
\Bigl(\sum_fi\bar q^f(0)\gamma_5 q^f(0)\Bigl)
\rangle \bar m^2.
\label{top_QCD}
\end{eqnarray}
This is an intriguing formula having the nonperturbative correlation in QCD between
the axial anomaly along with the $\theta$-vacuum and the chiral symmetry breaking
\footnote{
\textcolor{black}{
The relation only between the topological susceptibility and the quark condensates 
was already derived in 
Appendix of the literature \cite{Mao:2009sy}
by using the Ginzsparg-Wilson Dirac operator. 
Eq.(\ref{top_QCD}) has extended the existing formula to finite temperature, and by compensating the pseudoscalar susceptibility part (the last two terms). 
}
}
. 
Note that the topological susceptibility  
goes away, if either of quarks get massless. 
Thus the flavor-singlet nature of the axial
anomaly in QCD is surely reflected in the topological susceptibility in Eq.~(\ref{top_QCD})~\cite{Baluni:1978rf} (see also, e.g., \cite{Kim:2008hd}). 

\textcolor{black}{
As clearly shown in Appendix~\ref{chi5disc}, 
by using a couple of Ward identities, 
Eq.(\ref{top_QCD}) is reduced to the well-known formula relating 
the $\chi_{\rm top}$ to the disconnected part of the psuedoscalar susceptibility $\chi_{\rm 5,disc}$ including the flavor-singlet condition. 
In the reduced formula, the $\chi_{\rm 5,disc}$ implicitly includes the strange quark contribution in part, 
through the light-strange quark pseudoscalar susceptibility in such a way 
that $\chi_{\rm 5, disc} = m_s/(2m_l) \chi_P^{ls}$ (See Eq.(\ref{pseudo_disc})). 
In Eq.(\ref{top_QCD}) such a strange quark contribution has been made explicit.  
}

Even in the present linear sigma model  
the $\theta$-parameter must be entered 
only via the quark mass matrix ${\cal M}$, 
so that only the $V_{\rm SB}$ and the 
$V_{\rm SB-anom}$ parts include the $\theta$ dependence through ${\cal M}_{\theta}$ in Eq.(\ref{calM-theta}). 
In the present linear sigma model, thus, the $\theta$-dependent vacuum energy $V(\theta)$ is written as 
\begin{eqnarray}
V(\theta)&=&
\int_T d^4x\, V_{\rm SB}(\theta)+
\int_T d^4x\, V_{\rm SB-anom}(\theta)\nonumber\\
&=&\int_T d^4x
\Biggl(
-4cm_l\bar\Phi_1\cos \theta_l-2cm_s\bar\Phi_3 \cos \theta_s\nonumber\\
&&-8kcm_l\bar\Phi_1\bar\Phi_3\cos \theta_l-
4kcm_s(\bar\Phi_1)^2\cos\theta_s
\Biggl)
.
\end{eqnarray}
From Eq.~(\ref{topsus_dif}), the topological susceptibility of the linear sigma model
 reads
\begin{eqnarray}
\chi_{\rm top}
=
-\int_T d^4x\frac{\delta^2 V(\theta)}{\delta \theta(x)\delta \theta(0)}\Biggl |_{\theta=0}=
\left(\frac{2\langle \bar ll \rangle(T)}{m_l}+\frac{\langle \bar ss \rangle(T)}{m_s}
\right)\bar m^2,
\label{linear_chi}
\end{eqnarray}
where the light-quark condensate $\langle \bar ll \rangle$ and
the strange-quark condensate $\langle \bar ss \rangle$ should be evaluated from the linear sigma model analysis. 
This topological susceptibility 
is precisely in accordance with Eq.~(\ref{top_QCD}) at the leading order of $m_q$.
By taking the flavor universal limit $\langle \bar ll \rangle=\langle \bar ss \rangle\equiv \Sigma$, the expression of the topological susceptibility in Eq.~(\ref{linear_chi}) can be reduced to the Leutwyler-Smiluga (LS) relation~\cite{Leutwyler:1992yt}:  $\chi_{\rm top}\bigl|_{\rm LS}=\Sigma\bar m$, which is derived at the leading order in the ChPT.
It is remarkable that the $\chi_{\rm top}$ includes 
contributions from the anomaly-induced flavor breaking 
$k$-term, through the quark condensates in Eq.(\ref{condensates_tree}). 


\section{Formulation at finite temperature based on 
the CJT formalism}

When employing perturbative evaluation of thermal loop corrections in the linear sigma model at finite temperatures,   
we encounter the infrared divergence, because 
the meson loop corrections will be drastically 
enhanced to overwhelm the leading order terms 
due to the potentially small masses of pseudo-Nambu-Goldstone bosons. 
Therefore, some resummation scheme is required to avoid the infrared divergence.
Furthermore, performing the perturbative thermal loop calculations, some unphysical tachyonic mode would show up during a chiral phase transition. 
Thus, one may fail to implement the perturbative loop calculation in passing the phase boundary, 
which urges one to work on some nonperturbative 
analysis, to get reliable and physical results in 
the linear sigma model at finite temperatures.

In the present analysis, 
we shall employ the CJT formalism~\cite{Cornwall:1974vz}, which is  well-known and a powerful nonperturbative calculation tool to 
study the chiral phase transition based on the linear sigma model \cite{Lenaghan:2000ey,Roder:2003uz}.
The effective potential based on the CJT formalism is given as
\begin{eqnarray}
V_{\rm eff}[\alpha, S, P]&=&
V(\alpha)+\frac{1}{2}\int_k\left\{
[\ln  S^{-1}(k)]_{aa}+[\ln  P^{-1}(k)]_{aa}
\right\}\nonumber\\
&&
+\frac{1}{2}\int_k\left[
\bar S^{-1}_{ab}(k;\alpha) S_{ba}(k)+\bar P^{-1}_{ab}(k;\alpha) P_{ba}(k)
-2\delta_{ab}\delta_{ba}
\right]+V_2[\alpha, S, P],
\label{effectiveaction}
\end{eqnarray}
where $\alpha$ denotes a set of 
the expectation values of the scalar fields; 
$V(\alpha)$ is the tree-level potential given in Eq.~(\ref{TreePot}); 
$V_2[\alpha, S, P]$ includes contributions from  
the sum of all two-particle irreducible diagrams, 
in which all meson loop lines are drawn by 
the full (dressed) propagators, 
denoted as $S$ (for scalar mesons) and $P$ (pseudoscalar mesons).  
Here, we have used the shorthand notation
for the integration based on the imaginary time formalism,
\begin{eqnarray} 
\int_k f(k)=T\sum_{n=-\infty}^{\infty}\int \frac{d^3{\bm k}}{(2\pi)^3}f (2\pi i nT, {\bm k}).
\end{eqnarray}
The $\bar S^{-1}_{ab}(k;\alpha)$ and $\bar P^{-1}_{ab}(k;\alpha)$ are the tree-level propagators 
for scalar and pseudoscalar mesons respectively, 
\begin{eqnarray}
\bar S^{-1}_{ab}(k;\alpha)&=&-k^2\delta_{ab}+\left[m_S^2(\alpha)\right]_{ab},\nonumber\\
\bar P^{-1}_{ab}(k;\alpha)&=&-k^2\delta_{ab}+\left[m_P^2(\alpha)\right]_{ab},
\end{eqnarray}
where $\left[m_S^2(\alpha)\right]_{ab}$ and  $\left[m_S^2(\alpha)\right]_{ab}$
are the mass matrices of the scalar and pseudoscalar meson masses at the tree-level given in Eqs.~(\ref{TreeMassS}) and (\ref{TreeMassP}).
The expectation value of the one-point function 
$\bar \sigma$ and of the two-point functions denoted as $\bar {\cal S}$ and $\bar{\cal P}$ are determined 
from the stationary conditions,
\begin{eqnarray}
\frac{\delta V_{\rm eff}[\alpha, S, P]}{\delta \alpha_a}
\Biggl |_{\alpha=\bar\sigma, S=\bar{\cal S}, P=\bar{\cal P}}&=&0,\nonumber\\
\frac{\delta V_{\rm eff}[\alpha, S, P]}{\delta  S_{ab}}
\Biggl |_{\alpha=\bar\sigma, S=\bar {\cal S}, P=\bar {\cal P}}&=&0,\nonumber\\
\frac{\delta V_{\rm eff}[\alpha, S, P]}{\delta  P_{ab}}
\Biggl |_{\alpha=\bar\sigma, S=\bar {\cal S}, P=\bar {\cal P}}&=&0.
\label{stcond}
\end{eqnarray}
From latter two equations, 
the two-point functions can be expressed as 
\begin{eqnarray}
\bar {\cal S}_{ab}^{-1}(k)&=&\bar S^{-1}_{ab}(k;\bar \sigma)+\Sigma_{ab}(k),\nonumber\\
\bar {\cal P}_{ab}^{-1}(k)&=&\bar P^{-1}_{ab}(k;\bar \sigma)+\Pi_{ab}(k),
\label{TwoPonint}
\end{eqnarray}
where $\Sigma_{ab}(k)$ and $\Pi_{ab}(k)$ are the scalar- and pseudoscalar-self energy functions,
\begin{eqnarray}
\Sigma_{ab}(k)&=&2\frac{ \delta V_2[\alpha, S, P] }{\delta  S_{ba}}
\Biggl |_{\alpha=\bar\sigma, S=\bar {\cal S}, P=\bar {\cal P}}\nonumber\\
\Pi_{ab}(k)&=&2\frac{ \delta V_2[\alpha, S, P] }{\delta  P_{ba}}
\Biggl |_{\alpha=\bar\sigma, S=\bar {\cal S}, P=\bar {\cal P}}.
\label{SelfEnergy}
\end{eqnarray}
$\Sigma_{ab}$ and $\Pi_{ab}$ are functions of the $\bar {\cal S}_{ab}$ and $\bar{\cal P}_{ab}$, so that 
Eq.~(\ref{TwoPonint}) represents the the Schwinger-Dyson equations for the dressed propagators of the scalar and pseudoscalar mesons.

The $V_2[ \alpha, S, P]$ generically includes infinite number of diagrams, so 
an exact calculation is not practical. 
In the present study, we shall just pick up double-bubble diagrams, which is equivalent to the Hartree approximation. 
%
%
%
In this case, 
the effective potential for two-particle irreducible (2PI) diagrams $V_2$ is generated from the four-point interactions of the $\lambda_{1,2}$ terms in Eq.(\ref{mu_lam12}), and takes the form 
\begin{eqnarray}
V_2[ S, P]=
F_{abcd}
\left[\int_k{ S}_{ab}(k)\int_p{ S}_{cd}(p)+\int_k{ P}_{ab}(k)\int_p{ P}_{cd}(p)\right]
+2H_{abcd} \int_k{ S}_{ab}(k)\int_p{ P}_{cd}(p).
\end{eqnarray}
where
\begin{eqnarray}
F_{abcd}
&=&
\frac{\lambda_1}{8}(d_{abn}d_{ncd}+d_{adn}d_{nbc}+d_{acn}d_{nbd})
+\frac{\lambda_2}{4}(\delta_{ab}\delta_{cd}+\delta_{ad}\delta_{bc}+\delta_{ac}\delta_{bd}  )\nonumber\\
H_{abcd}&=&\frac{\lambda_1}{8}(d_{abn}d_{ncd}
+f_{acn}f_{nbd}+f_{bcn}f_{nad} )
+\frac{\lambda_2}{4}\delta_{ab}\delta_{cd}.
\end{eqnarray}
In the Hartree approximation, $V_2$ does not explicitly depend on $\alpha$, either are functions of $\alpha$.

The stationary conditions for the condensates $\bar \sigma_0$ and $\bar \sigma_8$ are then written as 
\begin{eqnarray}
c\sqrt{\frac{2}{3}}(2m_l+m_s)
&=&
\bar\sigma_0\left[
\mu^2+\frac{\lambda_1}{3}(\bar\sigma_0)^2+\lambda_2(\bar\sigma_0)^2-\frac{B}{\sqrt{6}}\bar\sigma_0
\right]
+(\bar\sigma_8)^2\left[
\lambda_1\bar\sigma_0+\lambda_2\bar\sigma_0-\frac{\lambda_1}{3\sqrt{2}}\bar\sigma_8+\frac{B}{2\sqrt{6}}
\right]\nonumber\\
&&
-kc\left[
\frac{m_l}{3}(8\bar\sigma_0-2{\sqrt{2}}\bar\sigma_8)
+\frac{m_s}{3}(4\bar\sigma_0+2\sqrt{2}\bar\sigma_8)
\right] \nonumber\\
&&-3G_{0bc}\left(\int_k{\bar {\cal S}}_{cb}(k)- \int_k{ \bar {\cal P}}_{cb}(k)\right)
+4F_{0bcd}\bar\sigma_d \int_k{\bar {\cal S}}_{cb}(k)
+4H_{0bcd}\bar\sigma_d \int_k{\bar {\cal P}}_{cb}(k)
,\nonumber\\
c\frac{2}{\sqrt{3}}(m_l-m_s)&=&
\bar\sigma_8\left[\mu^2+\frac{B}{\sqrt{6}}\bar\sigma_0
+\frac{B}{2\sqrt{3}}\bar\sigma_8+(\lambda_1+\lambda_2)\bar\sigma_0^2
-\frac{\lambda_1}{\sqrt 2}\bar\sigma_0\bar\sigma_8+
\left(\frac{\lambda_1}{2}+\lambda_2\right)(\bar\sigma_8)^2
\right]\nonumber\\
&&
-kc\left[
\frac{m_l}{3}(-2{\sqrt{2}}\bar\sigma_0-8\bar\sigma_8  )
+\frac{m_s}{3}(2\sqrt{2}\bar\sigma_0+2\bar\sigma_8  )
\right]
\nonumber\\
&&-3G_{8bc}\left(\int_k{\bar {\cal S}}_{cb}(k)- \int_k{\bar {\cal P}}_{cb}(k)\right)
+4F_{8bcd}\bar\sigma_d \int_k{\bar{\cal S}}_{cb}(k)
+4H_{8bcd}\bar\sigma_d \int_k{\bar{\cal P}}_{cb}(k),
\label{stforsigma}
\end{eqnarray}
where 
\begin{eqnarray}
G_{abc}=\frac{B}{6}\left[
d_{abc}-\frac{3}{2}(\delta_{a0}d_{0bc}+\delta_{b0}d_{a0c}+\delta_{c0}d_{ab0})+\frac{9}{2}d_{000}\delta_{a0}\delta_{b0}\delta_{c0}
\right].
\end{eqnarray} 
In the Hartree approximation, the self-energies 
$\Sigma_{ab}$ and $\Pi_{ab}$ included in 
the dressed propagators $\bar{\cal S}_{ab}$ and $\bar{\cal P}_{ab}$ (as in Eqs.(\ref{TwoPonint})) are independent of momenta. 
To solve the stationary conditions for $\bar\sigma$, $\bar {\cal S}$ and $\bar {\cal P}$ in Eq.~(\ref{stcond}), we  may therefore take an ansatz for 
the dressed propagator like 
\begin{eqnarray}
\bar {\cal S }^{-1}_{ab}(k)&=&-k^2\delta_{ab}+(M_S^2)_{ab},\nonumber\\
\bar {\cal P }^{-1}_{ab}(k)&=&-k^2\delta_{ab}+(M_P^2)_{ab}.
\end{eqnarray}
where $(M_S)_{ab}$ and $(M_P)_{ab}$ 
are dressed scalar and pseudoscalar masses in the matrix form, which are evaluated from Eqs.(\ref{TwoPonint}) and (\ref{SelfEnergy}) as 
\begin{eqnarray}
[M_{S}^2]_{ab}
&=&
[m_{S}^2(\bar \sigma)]_{ab}+
4F_{abcd}\int_k\bar {\cal S}_{cd}(k)+4H_{abcd} \int_k\bar {\cal P}_{cd}(k),\nonumber\\
{[}M_{P}^2]_{ab}
&=&
[m_{P}^2(\bar \sigma)]_{ab}+4F_{abcd}\int_k\bar {\cal P}_{cd}(k)+4H_{abcd} \int_k\bar {\cal S}_{cd}(k).
\label{stformass}
\end{eqnarray}
Note also that in the Hartree approximation, 
only the real part of the mass matrices are generated. 
The dressed-mass matrices can be diagonalized by  orthogonal transformations,
\begin{eqnarray}
{[}\tilde M_{S,P}^2]_{(i)}\delta_{ij}&=& \left(O'^{-1}_{S,P}\right)_{ia} [M_{S,P}^2]_{ab}\left(O'_{S,P}\right)_{bj},
\label{MixingMatT}
\end{eqnarray}
where $(\tilde M^2_{S,P})_{(i)}$ are mass eigenvalues for the dressed-masses. 
Note that in general  
the transformation matrices $O'_{S,P}$ are different from 
those at the tree-level (Eqs.~(\ref{transformT0S}) and (\ref{transformT0P})), 
due to the induced temperature dependence on the mixing angles $\theta_{S,P}$, so the mixing angles are computed as follows: 
\begin{eqnarray}
\theta_{S,P}=\frac{1}{2}\arctan\left[
\frac{2(M_{S,P}^2)_{08}}{(M_{S,P}^2)_{00}-(M_{S,P}^2)_{88}}
\right].
\label{mix_angle_T}
\end{eqnarray}
By using the transformation matrix $O'_{S,P}$ in Eq.~(\ref{MixingMatT}),
the dressed-propagators for $\bar{\cal S}$ and $\bar{\cal P}$ can be diagonalized as
\begin{eqnarray}
\tilde {\cal S}_{(i)}(k)\delta_{ij}&=&\left(O'^{-1}_{S}\right)_{ia}\bar{\cal S}_{ab}(k)\left(O'_{S}\right)_{bj},\nonumber\\
\tilde {\cal P}_{(i)}(k)\delta_{ij}&=&\left(O'^{-1}_{P}\right)_{ia}\bar{\cal P}_{ab}(k)\left(O'_{P}\right)_{bj},
\label{dressedP}
\end{eqnarray}
where $\tilde {\cal S}_{(i)}(k)$ and $\tilde {\cal P}_{(i)}(k)$ are 
the dressed propagators in the mass eigenbasis.

The loop integral for the dressed propagators in Eq.~(\ref{dressedP}) consists of
the vacuum-contribution part (zero-temperature part) plus the thermal-correction part.
Although the vacuum part has the ultraviolet (UV) divergences and  is subject to renormalization schemes, the thermal contribution to the loop integral is independent of UV divergence and renormalization. 
Actually, the qualitative results obtained based on the CJT analysis at finite temperatures 
are fairly insensitive to renormalization schemes, as was studied in~\cite{Lenaghan:1999si,Lenaghan:2000ey,Roder:2003uz}.
Thus, 
as far as qualitaiive features deduced from nonperturbative 
corrections are concerned, 
we may be almost free from the UV sensitivity, allowing to  
just work on the thermal corrections from the meson loops. 
Then the dressed propagators $\tilde {\cal S}_{(i)}(k)$ and $\tilde {\cal P}_{(i)}(k)$ are evaluated by only including 
the thermal-meson loop terms as  
\begin{eqnarray}
\int_k\tilde {\cal S}_{(i)}(k)&=&
\int\frac{d^3 {\bm k}}{(2\pi)^3}\frac{1}{\epsilon_{\bm k}[(\tilde M^2_S)_{(i)}]}
\left(
\exp\left\{\frac{ \epsilon_{\bm k}[(\tilde M^2_S)_{(i)}]}{T}\right\} -1
\right)^{-1},\nonumber\\
\int_k\tilde {\cal P}_{(i)}(k)&=&
\int\frac{d^3 {\bm k}}{(2\pi)^3}\frac{1}{\epsilon_{\bm k}[(\tilde M^2_P)_{(i)}]}
\left(
\exp\left\{\frac{ \epsilon_{\bm k}[(\tilde M^2_P)_{(i)}] }{T}\right\} -1
\right)^{-1},
\end{eqnarray}
where the $\epsilon_{\bm k}$ represents the relativistic energy of the mesons with momentum ${\bm k}$,
$\epsilon_{\bm k}[(\tilde M^2_{S(P)})_{(i)}] = \sqrt{{\bm k}^2+\left(\tilde M^2_{S(P)}\right)_{(i)}  }$.

The thermal-dependent 
quark condensates, $\langle \bar ll \rangle(T)$ and $\langle \bar ss \rangle(T)$, are evaluated as 
functions of the dressed propagators, and calculated 
as follows: 
\begin{eqnarray}
\langle \bar ll \rangle(T)&=&
\frac{\partial V_{\rm eff}[\alpha, S, P]}{\partial m_l}
\Biggl |_{\alpha=\bar\sigma, S=\bar{\cal S}, P=\bar{\cal P}}\nonumber\\
&=&\frac{\partial V(\alpha) }{\partial m_l}
\Biggl |_{\alpha=\bar\sigma}
+\frac{1}{2}\int_k\left[
\frac{\partial [m_S^2(\alpha)]_{ab}}{\partial m_l} S_{ba}(k)+\frac{\partial [m_P^2(\alpha)]_{ab}}{\partial m_l} P_{ba}(k)
\right]
\Biggl |_{\alpha=\bar\sigma, S=\bar{\cal S}, P=\bar{\cal P}}
\nonumber\\
&=&-2c(\bar \Phi_1(T) + 2k\bar \Phi_1(T) \bar\Phi_3(T))
+
\cdots,
\nonumber\\
\langle \bar ss \rangle(T)&=&
\frac{\partial V_{\rm eff}[\alpha, S, P]}{\partial m_s}
\Biggl |_{\alpha=\bar\sigma, S=\bar{\cal S}, P=\bar{\cal P}}\nonumber\\
&=&\frac{\partial V(\alpha) }{\partial m_l}
\Biggl |_{\alpha=\bar\sigma}
+\frac{1}{2}\int_k\left[
\frac{\partial [m_S^2(\alpha)]_{ab}}{\partial m_s} S_{ba}(k)+\frac{\partial [m_P^2(\alpha)]_{ab}}{\partial m_s} P_{ba}(k)
\right]
\Biggl |_{\alpha=\bar\sigma, S=\bar{\cal S}, P=\bar{\cal P}}
\nonumber\\
&=&-2c(\bar \Phi_3(T)+2k\bar \Phi_1^2(T))
+
\cdots
,
\label{quark_con}
\end{eqnarray}
where the thermal-meson loop corrections for quark condensates are included in terms denoted as 
$"\dots"$,
where are, in terms of the quark mass expansion, 
suppressed compared with the leading order terms in Eq.~(\ref{quark_con}). 
In numerically estimating the temperature dependence of the quark condensates,
we will pick up only the leading terms in Eq.~(\ref{quark_con}).
Since 
the $\bar \Phi_1$ and $\bar \Phi_3$ 
are determined by solving the stationary conditions in Eqs.~(\ref{stforsigma}) and (\ref{stformass}), so 
that the quark condensates nonperturbatively get  thermal effects. 
Of interest is that the quark condensates explicitly have a remnant of the axial-anomaly induced-flavor breaking, what we call the $k$-term, $V_{\rm SB-anom}$, as seen in Eq.~(\ref{quark_con}). Thus, the axial-anomaly induced-flavor breaking would be expected to directly contribute to the  
chiral restoration phenomena at finite temperatures.

\section{Numerical analysis on nonperturbative flavor breaking 
at chiral crossover criticality}

We are now ready to study how the nonperturbative 
flavor breaking can be generated at finite 
temperatures across the chiral critical phenomena, 
such as the chiral crossover. 
To this end, 
we take the following values for the model 
parameters as inputs 
\begin{eqnarray}
\mu^2&=&1.02\times 10^4\,{\rm MeV}^2,\;\;\;
\lambda_1=11.8,\;\;\;
\lambda_2=20.4,\nonumber\\
c m_l&=& 6.11\times 10^5\,{\rm MeV}^3,\;\;\;
cm_s=198\times 10^5\,{\rm MeV}^3\nonumber\\
B&=&3.85\times 10^3\,{\rm MeV},\;\;\;
k=3.40\,{\rm GeV}^{-1}, 
\end{eqnarray} 
with which the present model at vacuum 
well reproduces the scalar and pseudoscalar meson 
spectra, including the inverse mass hierarchy for 
scalar mesons ligher than 1 GeV~\cite{Kuroda:2019jzm}:
From Eqs.~(\ref{TreeMassS}) and (\ref{TreeMassP}) with the input parameters as above,  
the meson masses in the vacuum (at zero temperature) are estimated to be 
\begin{eqnarray}
m[f_0(500)]&=&672.4\,{\rm MeV},\;\;\;
m[f_0(980)]=990.4\,{\rm MeV},\;\;\;
m_{a_0}=937.6\,{\rm MeV},\;\;\;
m_{\kappa}=863.4\,{\rm MeV},\nonumber\\
m_{\eta'}&=&958.2\,{\rm MeV},\;\;\;
m_\eta=552.9\,{\rm MeV},\;\;\;
m_{\pi}=137.9\,{\rm MeV},\;\;\;
m_{K}=494.1\,{\rm MeV},
\end{eqnarray}
which indeed shows good agreement with the experimental 
values~\cite{Tanabashi:2018oca}. 

In addition, 
we predict the the topological susceptibility to be  
\begin{align} 
\chi_{\rm top}(T=0)\simeq0.0263/{\rm fm}^4 
\qquad {\rm or} \qquad  
\left(\chi_{\rm top}(T=0)\right)^{1/4}\simeq 79.4\, 
{\rm MeV}. 
\end{align} 
Actually, the recent lattice QCD data, with 2 + 1 (+1) flavors having a physical pion mass and the continuum limit being taken, predicts the topological susceptibility at the zero temperature as $\chi_{\rm top}(T=0)=0.019(9)/{\rm fm}^4$~\cite{Bonati:2015vqz}, and $\chi_{\rm top}(T=0)=0.0245(24)_{\rm stat}(03)_{\rm flow}(12)_{\rm cont}/{\rm fm}^4$~\cite{Borsanyi:2016ksw}. 
For the latter the first error is statistical, the second error is systematic error and the third error comes from changing the upper limit of the lattice spacing range in the fit. 
On the lattice, the topological susceptibility is defined through
a certain smoothing process to reduce UV sensitivity of
the topological charge operator, which brings a systematic error.
Besides, sampling among different topological sectors is a nontrivial task for lattice QCD simulations~\cite{Vicari:2008jw,Luscher:2010iy,Alexandrou:2015yba,Alexandrou:2017hqw,Bonati:2018blm}.  
Although their central values do not agree each other, from a conservative point of view, we may say that the
difference between them is interpreted as a systematic error from the individual lattice QCD calculation.


\subsection{Chiral order parameters}

\begin{figure}[H]
  \begin{center}
   \includegraphics[width=10cm]{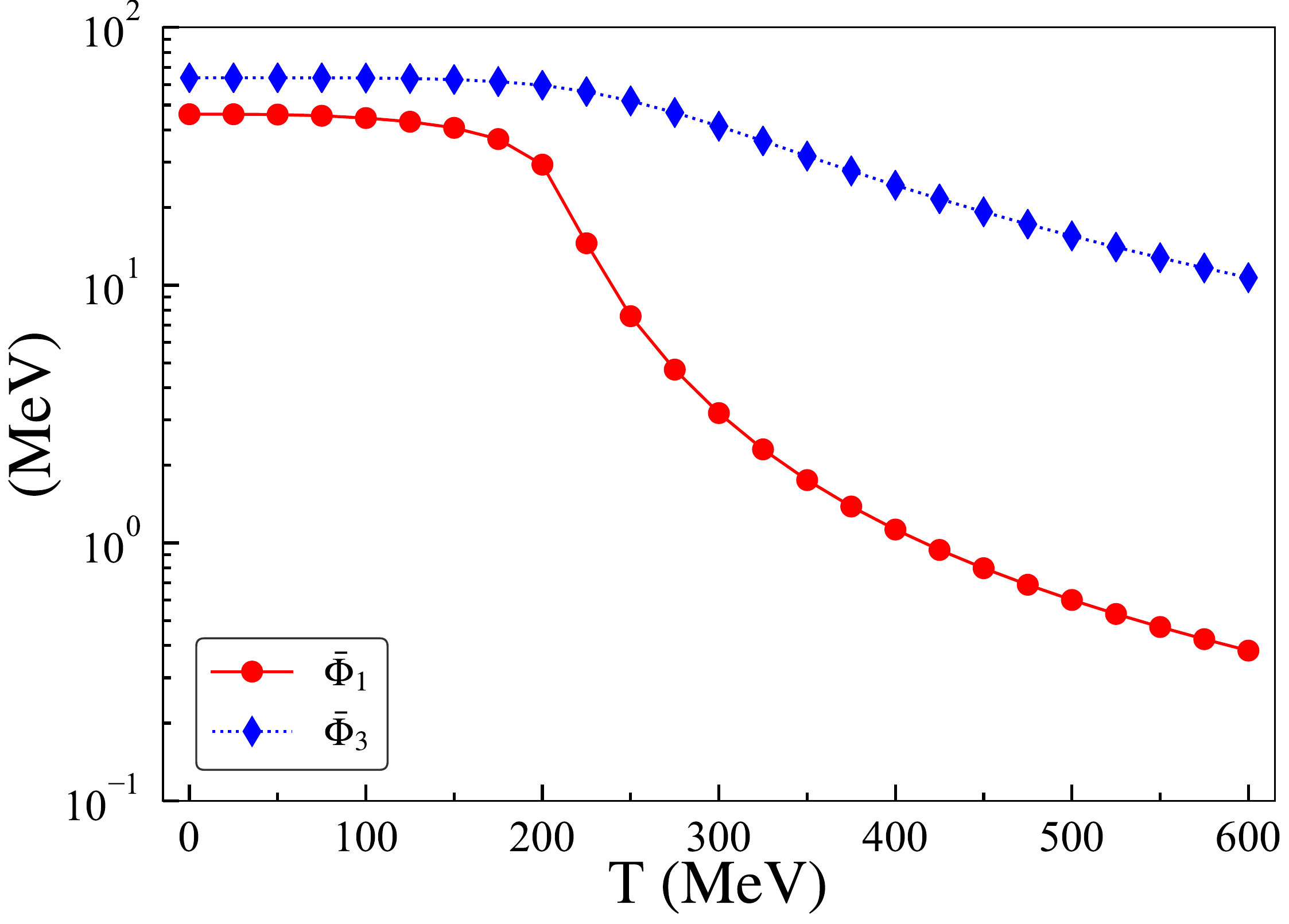}
  \end{center} 
 \caption{
The $\bar \Phi_{1}$ and $\bar \Phi_{3}$ as a 
function of the temperature.
}  
 \label{condensation_fig}
\end{figure}

In Fig.~\ref{condensation_fig},
we first show a plot on the scalar condensates $\bar \Phi_{1,3}$ as a function of temperature,
which are evaluated by solving the stationary conditions for the CJT effective potential in Eq.~(\ref{stcond})~\footnote{
At the vacuum $T=0$, the scalar condensates are well degenerated,
$\bar \Phi_{3}(T=0)/\bar \Phi_{1}(T=0) \simeq1.38$.
It indicates that the three-flavor linear sigma model approximately has the $SU(3)$-flavor symmetry 
at the vacuum.}. 
The figure tells us that 
the $\bar\Phi_1$ becomes smaller rapidly than the $\bar\Phi_3$ does. 
This tendency 
implies that the two-flavor chiral symmetry, 
controlled by the $\Phi_1$,  
can be restored faster than that for the heavier strange quark, by the $\Phi_3$. 
However, it looks like that 
the chiral symmetry is not exactly restored, 
so it may be sort of crossover phenomenon. 
The similar scaling property of scalar condensates with respect to the temperature has been observed in other three-flavor models based on the CJT  formalism~\cite{Lenaghan:2000ey,Roder:2003uz}, in which the axial-anomaly induced-flavor breaking, the $k$-term in $V_{\rm SB-anom}$, is not incorporated.

\begin{figure}[H]
  \begin{center}
   \includegraphics[width=10cm]{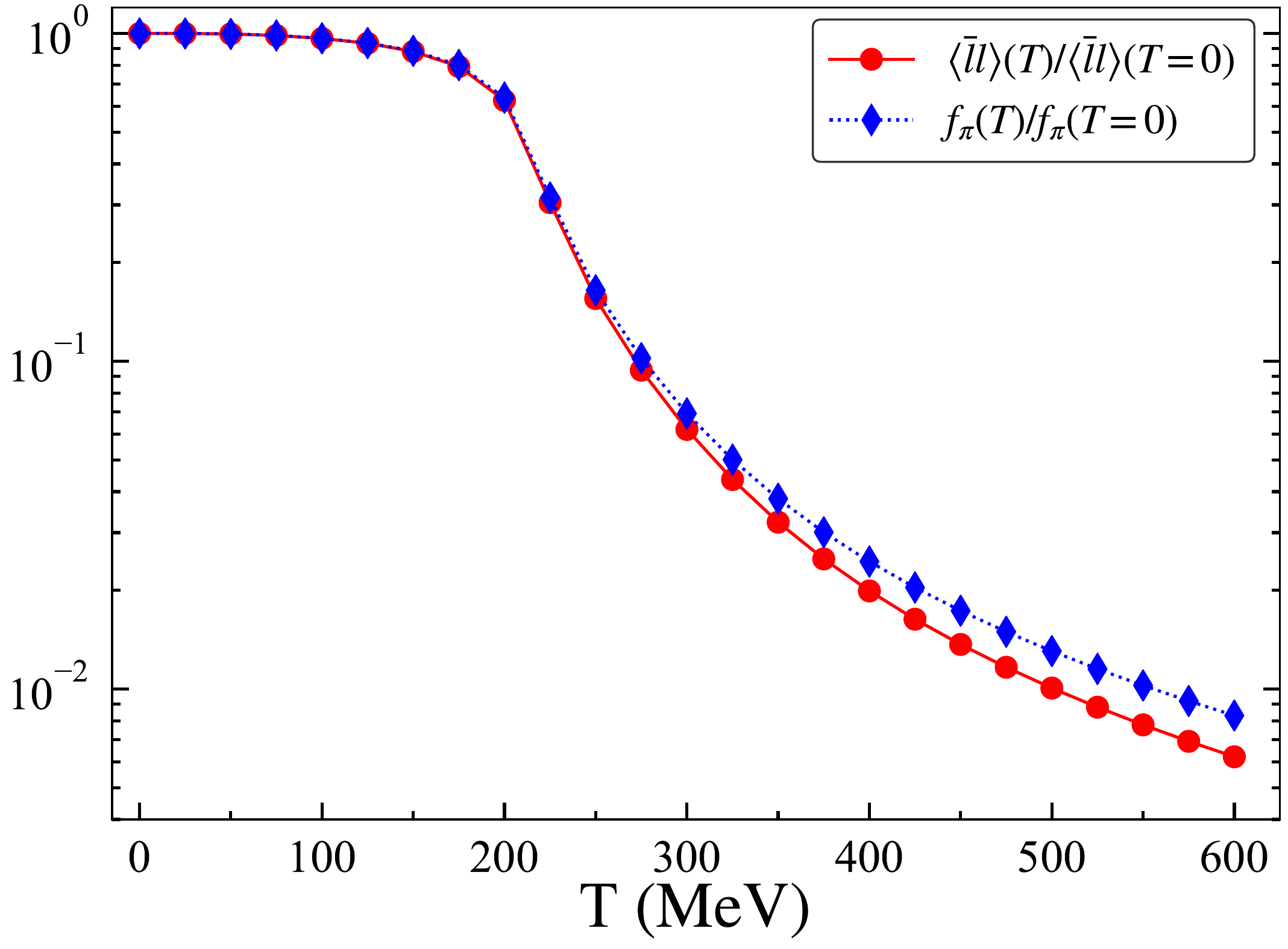}
  \end{center} 
 \caption{ 
 The temperature dependence of the light quark condensates and the pion decay constant, normalized to 
 those vacuum values. 
}  
 \label{quark_con_plot}
\end{figure}

We next discuss thermal effects on the quark condensates $\langle \bar qq \rangle(T)$ in Eq.~(\ref{quark_con}). 
Fig.~\ref{quark_con_plot} displays 
the temperature dependence of the light quark condensate $\langle \bar ll\rangle(T)/\langle \bar ll\rangle(T=0)$. The figure shows that the light quark condensate 
does not exactly reach zero, though being monotonically damping as $T$ gets larger. 
This implies that the chiral phase transition does not happen, instead, the theory undergoes a chiral crossover, as was indicated in the scaling of scalar condensates above. 
This crossover phenomenon is in a qualitative sense consistent with the current result of the lattice QCD with $2+1$ flavors~\cite{Aoki:2006we,Borsanyi:2011bn,Ding:2015ona,Bazavov:2018mes,Ding:2020rtq}. 
It it interesting to note that the chiral crossover 
has been observed 
even in the absence of the axial-anomaly induced-flavor breaking term $V_{\rm SB-anom}$, so, in this sense, the $k$ term is irrelevant to realize 
the chiral crossover.

From Fig.~\ref{quark_con_plot}, we may estimate the pseudo-critical temperature of the chiral crossover, simply by identifying it as a temperature $T_{pc}^*$, where  
$d^2 \langle \bar{l}l \rangle(T)/dT^2|_{T=T_{pc}^*}=0$. 
We then find $T^*_{pc}\simeq 215$ MeV. 
Note that the definition of our $T_{pc}^*$ 
is different from the 
lattice QCD's yielding the pseudo-critical temperature $T_{pc}\simeq 155$ MeV
and has been estimated to be  
larger~\cite{Aoki:2006we,Borsanyi:2011bn,Ding:2015ona,Bazavov:2018mes,Ding:2020rtq}: 
By construction, 
the present chiral effective model can only include 
operators with single current quark mass matrix ${\cal M}$, so we cannot 
evaluate the chiral susceptibility, through which the pseudo-critical temperature in the lattice simulation is defined. 
However the temperature at which the maximum of the chiral susceptibility should reach is usually compatible with the inflection point of the chiral condensate in lattice QCD data, i.e. the point where 
$d^2 \langle \bar{l}l \rangle(T)/dT^2|_{T=T_{pc}}=0$. 
So, we may naively have quantitative comparison between $T_{pc}$ and $T_{pc}^*$, 
to find about 30\% deviation. 
We will come back to this quantitative discrepancy in 
the present analysis, in the later section.

To make manifest how the axial-anomaly induced-flavor breaking affects the chiral crossover phenomenon, 
we compare the ratio 
$\langle \bar ll\rangle(T)/\langle \bar ll\rangle(T=0)$ with the pion decay constant normalized to the vacuum value $f_\pi(T)/f_\pi(T=0)$, which corresponds to the temperature dependence of the light quark condensates in the absence of the axial-anomaly induced-flavor breaking term. 
Fig.~\ref{quark_con_plot} shows that the $\langle \bar ll \rangle(T)/\langle \bar ll \rangle(T)$ drops somewhat rapidly than the 
$f_\pi(T)/f_\pi(T=0)$. 
This implies that the anomaly-induced flavor breaking acts as a catalyzer toward the chiral restoration. 


\begin{figure}[H]
  \begin{center}
   \includegraphics[width=10cm]{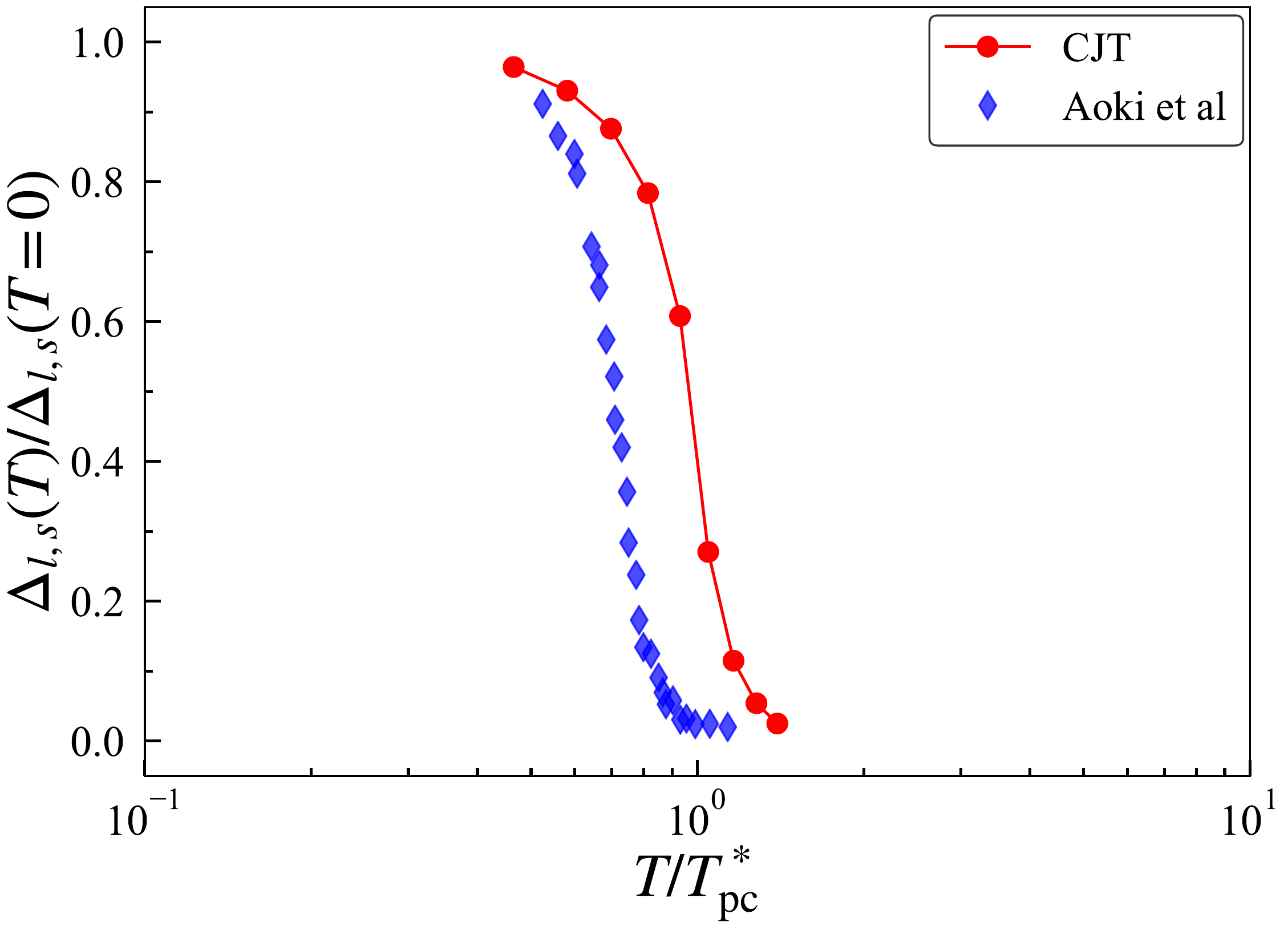}
  \end{center} 
 \caption{ 
The comparison of the subtracted chiral condensate $\Delta_{l,s}(T)$ with the lattice QCD data~\cite{Aoki:2009sc}.
}  
 \label{subcon}
\end{figure}

The chiral condensate has the UV divergence and is need to be renormalized to get the finite quantity. To eliminate the quadratic divergences tagged with the quark mass in the chiral condensates, we use the subtracted chiral condensate 
$\Delta_{l,s}(T)=\langle \bar ll \rangle -\frac{2m_l}{m_s}\langle \bar ss \rangle$ 
as the order parameter of the chiral phase transition:
Fig.~\ref{subcon} shows the temperature dependence of the subtracted chiral condensate based on the CJT analysis, in comparison with the lattice QCD observation~\cite{Aoki:2009sc}.
Although the deviation from the lattice QCD data~\cite{Aoki:2009sc} is read as about $30\%$
around the pseudo-critical temperature regions $T\simeq T_{pc}^*$, the CJT analysis qualitatively supplies the chiral crossover as observed in the lattice QCD observation~\cite{Aoki:2009sc}.

\begin{figure}[H]
\begin{center}
   \includegraphics[width=7.0cm]{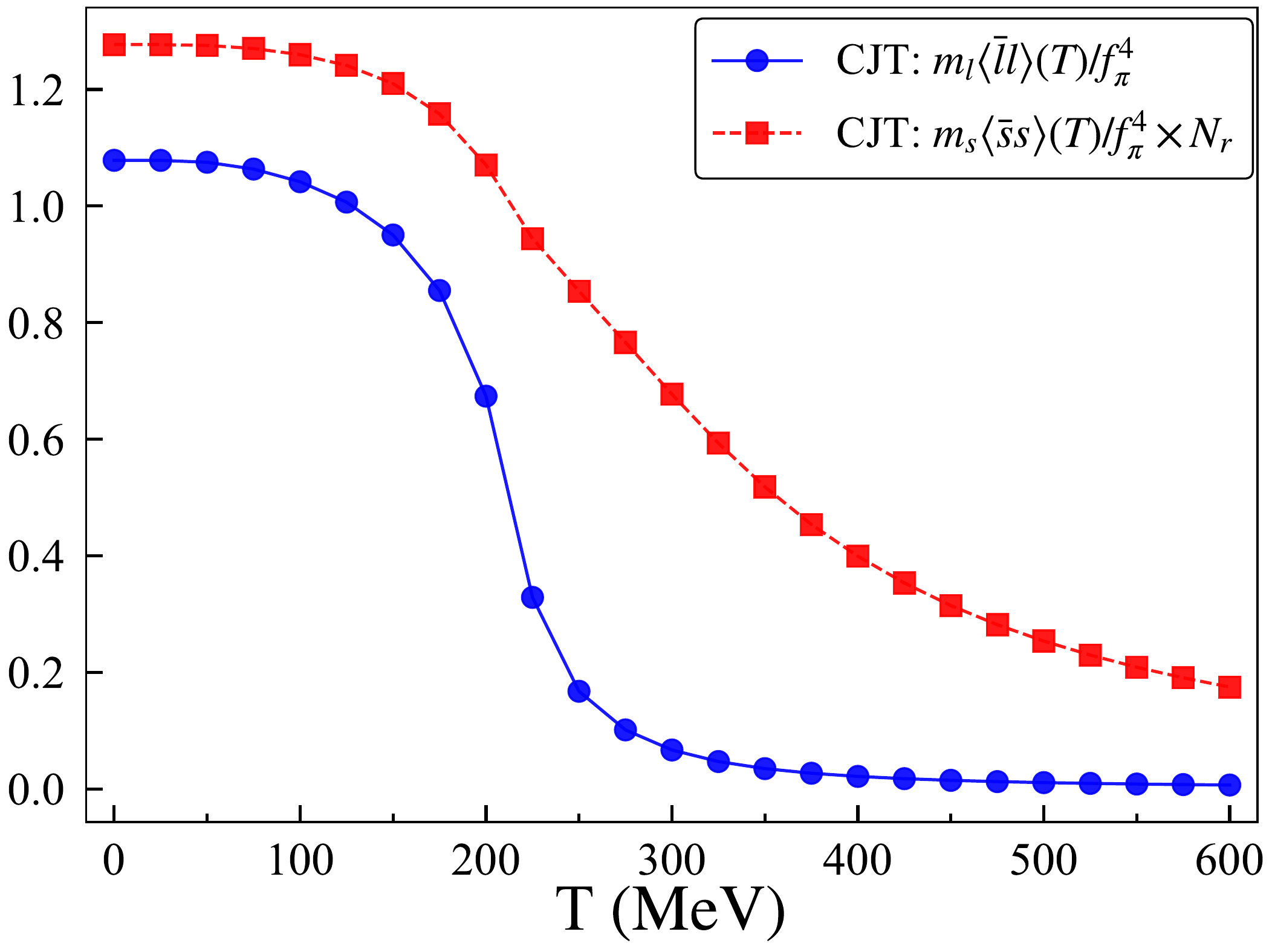}
    \subfigure{(a)}
   \includegraphics[width=7.0cm]{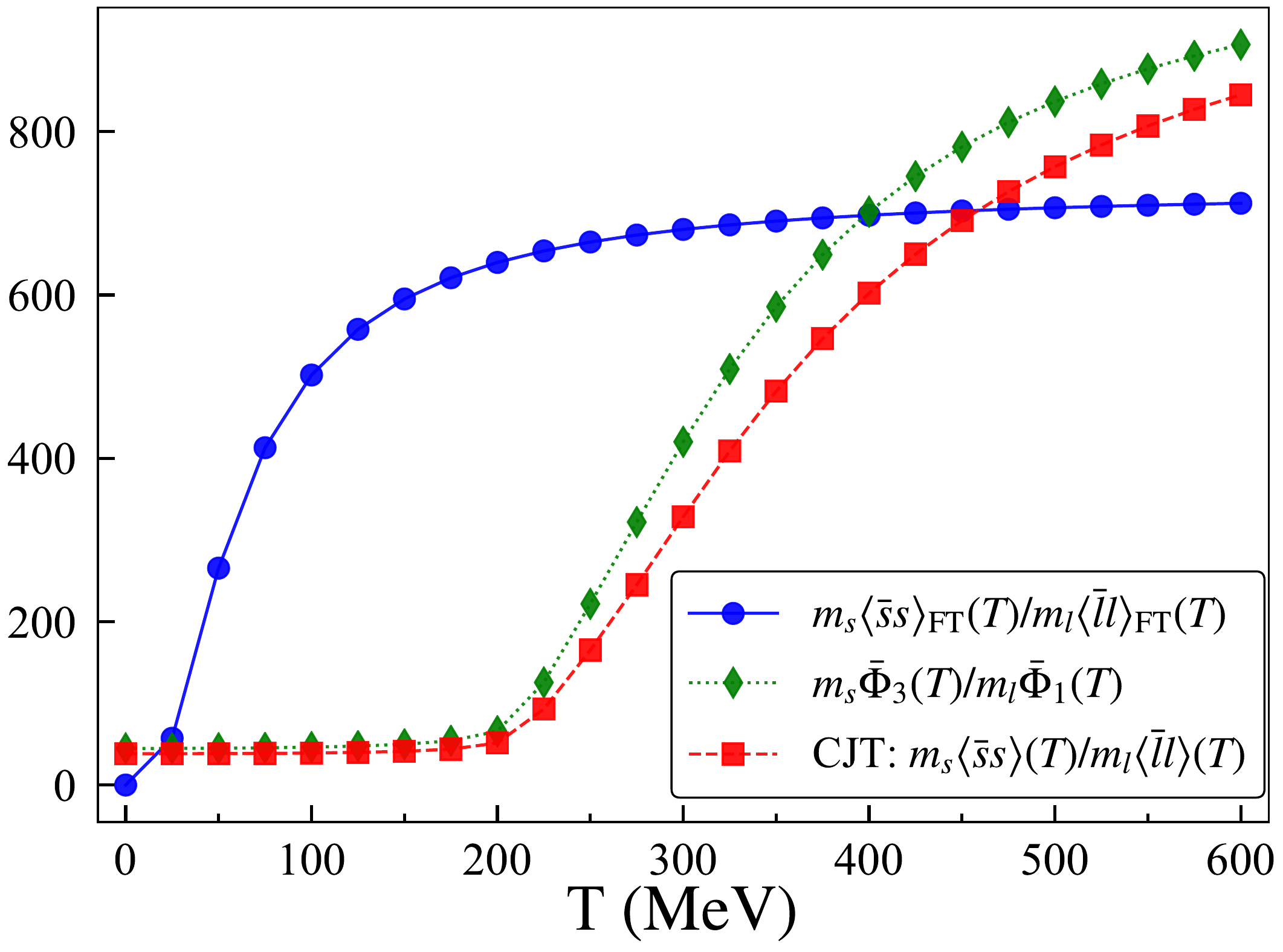}
    \subfigure{(b)}
  \end{center}
 \caption{
 Plots on the temperature dependence of the renormalization-scale invariant 
 quark-condensates $m_s \langle \bar ss \rangle(T)$ 
 and $m_l\langle \bar ll \rangle(T) $ normalized to $f_\pi^4$, 
 individually in panel (a) and 
 via its ratio together with comparison with $m_s \bar \Phi_3(T)/m_l \bar \Phi_1(T)$
and 
$m_s \langle \bar ss \rangle_{\rm FT}(T)/m_l \langle \bar ll \rangle_{\rm FT}(T)$ obtained in the free theory of quarks, in panel (b).  
 In panel (a) the $m_s \langle \bar ss \rangle(T)/f_\pi^4$ has been multiplied by 
 a factor of 
 $N_r = 1/32.4$, corresponding to the value of $m_l/m_s$ estimated from the present 
 linear sigma model. 
 }
\label{RGI_qcon}
\end{figure}

We now show 
the significance of the flavor breaking between the light and strange quark condensates at around 
the chiral crossover. 
Note again that quark condensates themselves potentially include 
UV divergences, hence dependence on the renormalization scales, 
so would be unreasonable to merely compare those for light and strange 
flavors, in light of the lattice observation. 
In Fig.~\ref{RGI_qcon} we thus plot the renormalization-scale invariant combinations 
$m_s \langle\bar s s \rangle(T)$ 
and $m_l \langle \bar ll \rangle(T)$ normalized to $f_\pi^4$ (panel (a)) 
and its ratio $m_s \langle\bar s s \rangle(T)/m_l \langle\bar l l \rangle(T)$ 
(panel (b)), as a function of the temperature. 
At the zero temperature, the ratio is computed as $m_s \langle \bar ss \rangle(T=0)/m_l \langle \bar ll \rangle(T=0)\simeq (32.4) \times (1.18) \simeq 38$ 
(with $m_s/m_l \simeq 32.4$), which is in good agreement with the recent lattice simulation (at a physical pion mass and the renormalization scale of 2 GeV), $m_s \langle \bar{s}s \rangle(T=0)/m_l \langle \bar{l}l \rangle(T=0) = (32.4) \times (1.08 \pm 0.16)$~\cite{McNeile:2012xh}. 
Therefore, 
the $SU(3)_V$ flavor symmetry between the light quark and the strange quark condensate is approximately preserved at the zero temperature.
Moving on to the finite temperature environment, 
the ratio $m_s \langle \bar ss \rangle(T)/m_l \langle \bar ll \rangle(T)$ is drastically enhanced and monotonically increased around the pseudo-critical temperature $T_{pc}^* \simeq 215$ MeV, so that the violation of flavor symmetry between the light quark and the strange quark condensates gets more eminent over the chiral criticality~\cite{Kawaguchi:2020kdl}. 
This dramatic enhancement has been also observed in several analyses on hot lattice QCD~\cite{Cheng:2007jq,Nicola:2016jlj}. 
Furthermore, to extract the contribution of $V_{\rm SB-anom}$ to the flavor breaking, we compare the $m_s \langle \bar ss \rangle(T)/m_l \langle \bar ll \rangle(T)$ with the $m_s \bar \Phi_3(T)/m_l \bar \Phi_1(T)$ which is identical to 
the one without the anomaly-induced flavor breaking effect.
Fig.~\ref{RGI_qcon} shows that the axial-anomaly induced-flavor breaking term plays a role to reduce the magnitude of the flavor breaking.

To clarify that 
the flavor breaking certainly comes from the nonperturbative 
thermal contribution, 
in Fig.~\ref{RGI_qcon} we also compare the $m_s \langle \bar ss \rangle(T)/m_l \langle \bar ll \rangle(T) $ with the one corresponding to the free theory (FT) of quarks, in which the quarks behave non-interacting free-particles. The ratio of free-quark condensates $m_s \langle\bar ss \rangle_{\rm FT}(T)/m_l\langle\bar l l \rangle_{\rm FT}(T)$ is perturbatively obtained from the thermal one-loop calculation.
We provide the detailed expression of quark condensates in the FT in the Appendix B. 
For the low temperature regions where $T<m_s \sim 100$ MeV, 
the FT-quark condensate ratio $m_s \langle\bar ss \rangle_{\rm FT}(T)/m_l\langle\bar l l \rangle_{\rm FT}(T)$ keeps the value below $(m_s/m_l)^2 \simeq (27)^2=729$, because of the naive Boltzmann suppression for the strange quark contribution. 
After passing $T\simeq m_s\sim 100$ MeV, the $m_s \langle\bar ss \rangle_{\rm FT}(T)/m_l \langle\bar l l \rangle_{\rm FT}(T)$ asymptotically approaches $(m_s/m_l)^2 = 729$, which merely reflects the trivial and overall flavor-breaking just by quark masses: $m_q \langle \bar{q}q \rangle \sim m_q^2 T^2$. 
In contrast, in whole low-temperature regions  $T<T_{pc}^* \simeq 215$ MeV the CJT analysis exhibits a gigantic suppression 
for the quark-condensate ratio more than the Boltzmann suppression, to respect the flavor symmetry, i.e., the vacuum value $\sim (m_s/m_l) \times 1$. 
This flavor symmetric behavior is consistent with the conventional three-flavor ChPT observation, 
in which the vectorial $SU(3)$ flavor symmetry cannot explicitly be violated at the leading order of the chiral expansion based on the nonlinear-sigma model setup, 
so that 
$m_s \langle\bar ss \rangle_{\rm ChPT}(T)/m_l \langle\bar l l \rangle_{\rm ChPT}(T)\simeq (m_s/m_l) \times 1$ even including 
the next-to leading order corrections. 
As the temperature further increases, the CJT analysis nonperturbatively undergoes the chiral crossover around the pseudo-critical temperature $T_{pc}^* \simeq 215$ MeV (see Figs.~\ref{condensation_fig}, ~\ref{quark_con_plot} and panel (a) of \ref{RGI_qcon}), where the light quark condensate $m_l \langle \bar ll \rangle(T)$ starts to drop more efficiently than the strange quark condensate $m_s \langle \bar ss \rangle(T)$. 
Consequently, the quark-condensate ratio rapidly starts to grow from $T\simeq T_{pc}^*$.
This is a nontrivial flavor breaking, 
essentially different from the trivial FT flavor violation 
just by $m_s/m_l$. 
In the end, at around $T\simeq 600$ MeV, the quark-condensate ratio asymptotically merges with the FT yielding the trivial-flavor breaking value $(m_s/m_l)^2$. 
This would imply that the present linear sigma 
model would converge to an ideal-quark gas picture 
consistently with the asymptotic free nature of 
the underlying QCD.
Thus we can conclude that what we call a nonperturbative flavor breaking is certainly a nonperturbative output in association with the characteristic chiral crossover phenomenon.

\subsection{Meson spectral properties} 
\label{mix_angle}

The feature of the chiral crossovver criticality should be reflected in the meson dynamics. 
In particular, the mixing of singlet and octet states are  strongly correlated with the chiral symmetry breaking.
In this subsection, we first discuss the mixing structure of constituents in the $f_0(500)$ and $f_0(980)$ ($\eta'$ and $\eta$) via the temperature dependent mixing angles 
$\theta_{S,P}$ in Eq.~(\ref{mix_angle_T}), across the 
crossover criticality.

\begin{figure}[H]
  \begin{center}
   \includegraphics[width=10cm]{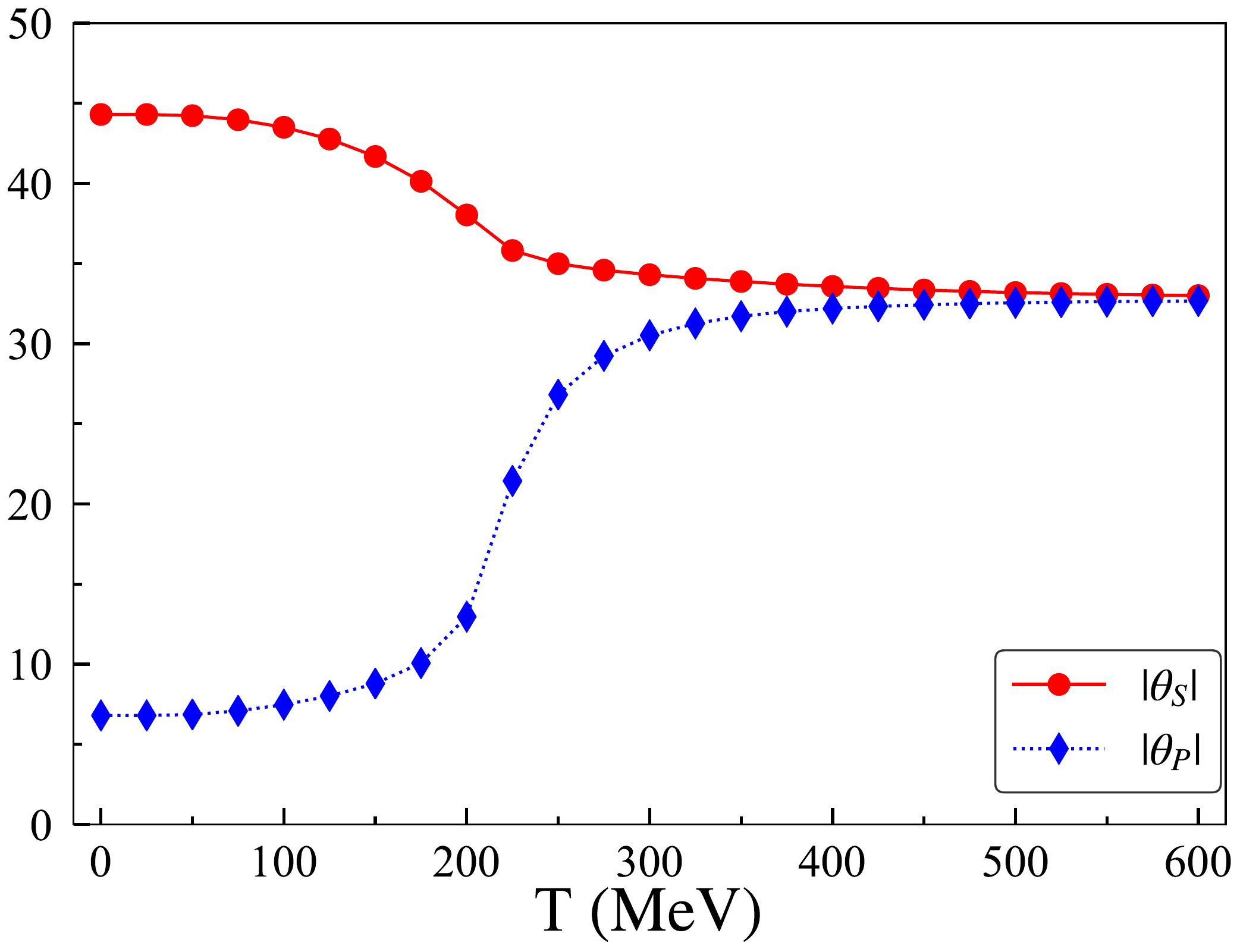}
  \end{center} 
 \caption{ 
The mixing angles of the scalar ($\sigma_0, \sigma_8$) and the pseudoscalars 
$(\pi_0, \pi_8)$ as a function of temperature.
}  
 \label{mixing_plot}
\end{figure}

Fig.~\ref{mixing_plot} shows the $\theta_{S,P}$ as a function of temperature. When the temperature is increased up to around the the pseudo-critical temperature $T^*_{pc}\simeq 215$ MeV, the $\theta_{S}$ becomes smaller. On the other hand, the $\theta_{P}$ dramatically grows. After arriving at around $T^*_{pc}$, the mixing angles $\theta_{S,P}$ converge gradually to the ideal mixing angle $\theta_{S,P}=\arcsin(1/\sqrt{3}) \simeq 35.3^\circ$, where the mass eigenstates completely 
separate into the strange-quark and non-strange quark scalars~\footnote{  
Indeed, at higher temperatures around $T\simeq 600$ MeV, where the ideal mixing is almost perfectly realized, 
the $f_0(500)$ and $\eta'$ meson are completely made of  the fluctuating modes from $\bar\Phi_1$ (non-strange, or light-quark components), while 
the $f_0(980)$ and the $\eta$ meson are 
the fluctuating modes from $\bar\Phi_3$ (strange-quark components).}.   
Therefore, this implies that above $T_{pc}^*$
the $f_0(500)$ and $\eta'$ mesons are identified as  
light-quark ($l\bar l$) scalar sates, and the $f_0(980)$ and $\eta$ mesons as $s\bar s$ states. 
Actually, other three-flavor model analyses based on the CJT formalism~\cite{Lenaghan:2000ey} have also predicted the same critical phenomenon for the mixing structure, where 
the anomaly-related flavor breaking term $V_{\rm SB-anom}$ is not involved. 
In that sense, we may conclude that the axial-anomaly induced-flavor breaking is not essential so much for 
the meson mixing structure in relation to 
the critical phenomenon.

\begin{figure}[H]
\begin{tabular}{cc}
 \begin{minipage}{0.5\hsize}
  \begin{center}
   \includegraphics[width=7.4cm]{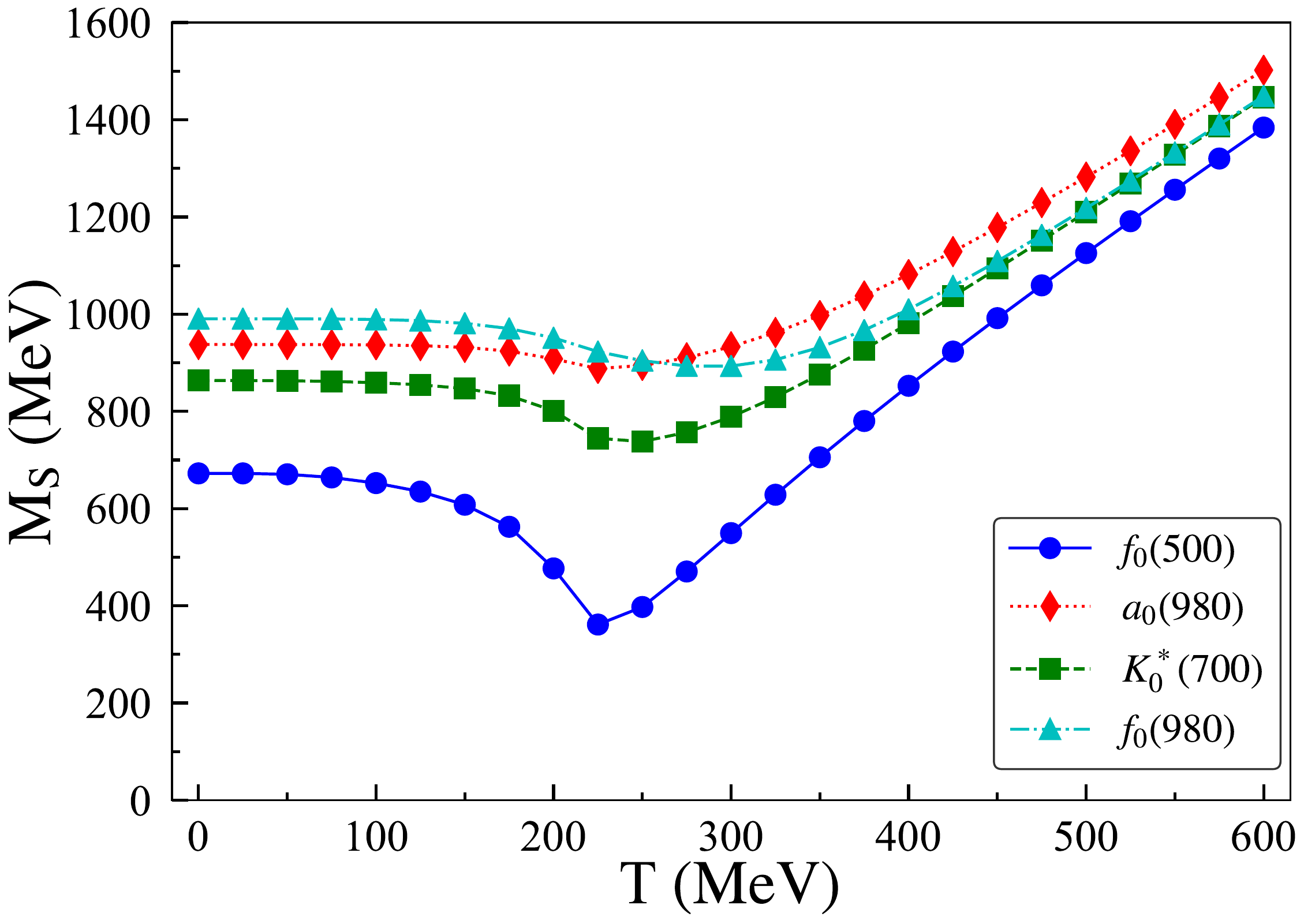}
    \subfigure{(a)}
  \end{center}
 \end{minipage}%
 \begin{minipage}{0.5\hsize}
  \begin{center}
   \includegraphics[width=7.4cm]{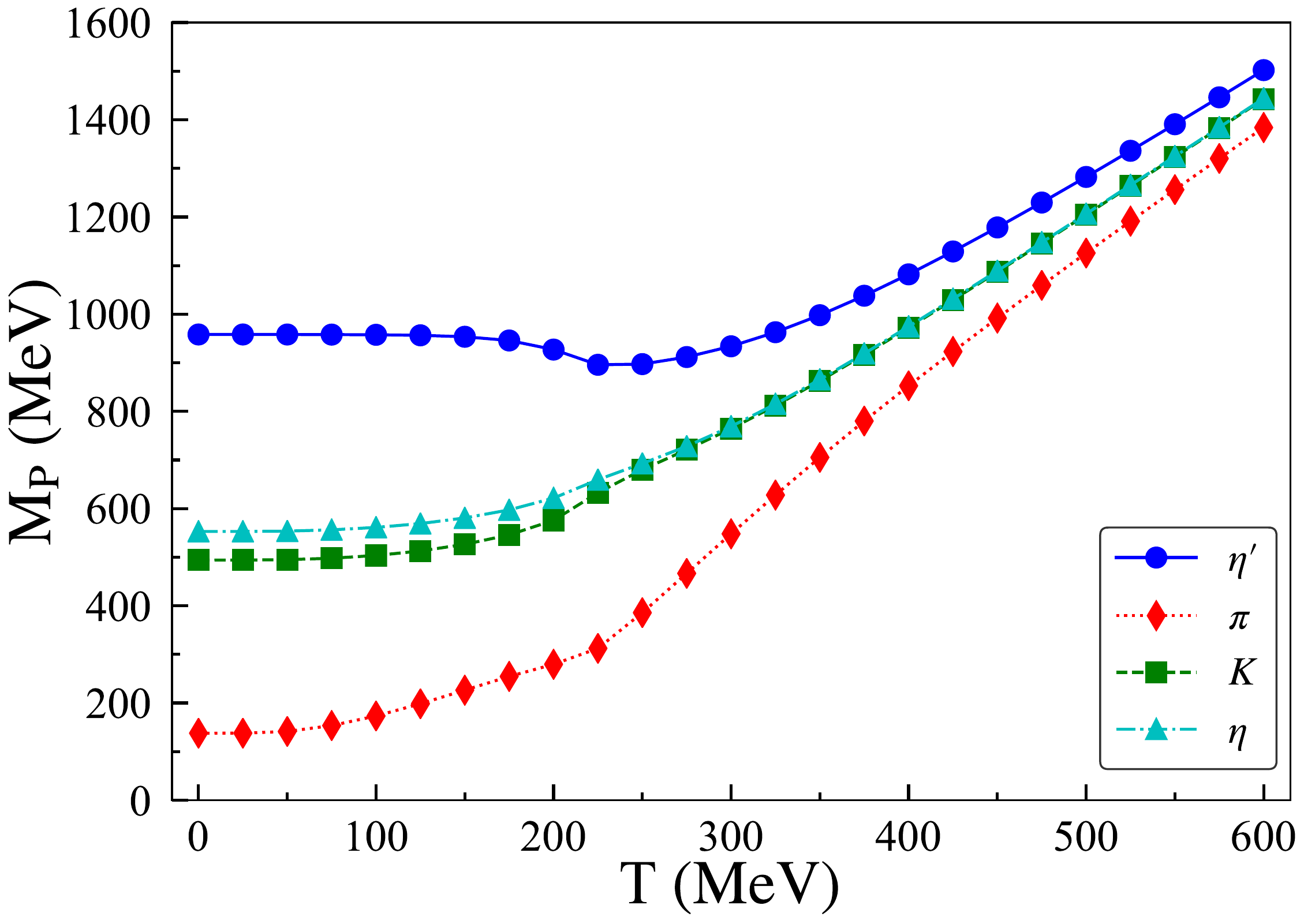}
    \subfigure{(b)}
  \end{center}
 \end{minipage}
 \end{tabular}
 \caption{ The thermal effects on the dressed masses   
for (a) scalar mesons, and (b) pseudoscalar mesons.
}  
 \label{ploteffmass}
\end{figure}

Second, we focus on the criticality effect on 
the meson masses. 
It turns out that a couple of interesting 
consequences of the chiral crossover 
emerge in meson spectroscopy.
In Fig.~\ref{ploteffmass}, we show the temperature dependence of the dressed masses for scalar mesons and the pseudoscalar mesons.
When the temperature is increased up to around the pseudo-critical temperature $T^*_{pc}$, the dressed-scalar masses drop. 
In particular, the dropping rate for the 
$f_0(500)$ meson gets prominent. 
As for the pseudoscalar masses, the dressed mass of $\eta'$ slightly decreases below the pseudo-critical temperature  $T^*_{pc}$, while the other pseudoscalar masses grow. 
Above $T_{pc}^*$, 
especially at $T\geq 300\,{\rm MeV}$, 
all of the dressed-scalar and -pseudoscalar masses monotonically increase and tend to degenerate.
We shall discuss more details on these scaling properties below.  

It would be interesting to compare the thermal masses in Fig.~\ref{ploteffmass} with the screening masses observed on the lattice QCD~\cite{Bazavov:2019www}. 
We find that the mass scaling properties of the pseudoscalars qualitatively agree with the lattice data. 
Regarding the scalar mesons, accurate lattice studies on identifying the bound state channels currently involve some complicated issues. 
Therefore, direct comparison to our predictions is not straightforward, which would be possibly resolved by the future studies.

\begin{figure}[H]
\begin{tabular}{cc}
 \begin{minipage}{0.5\hsize}
  \begin{center}
   \includegraphics[width=7.4cm]{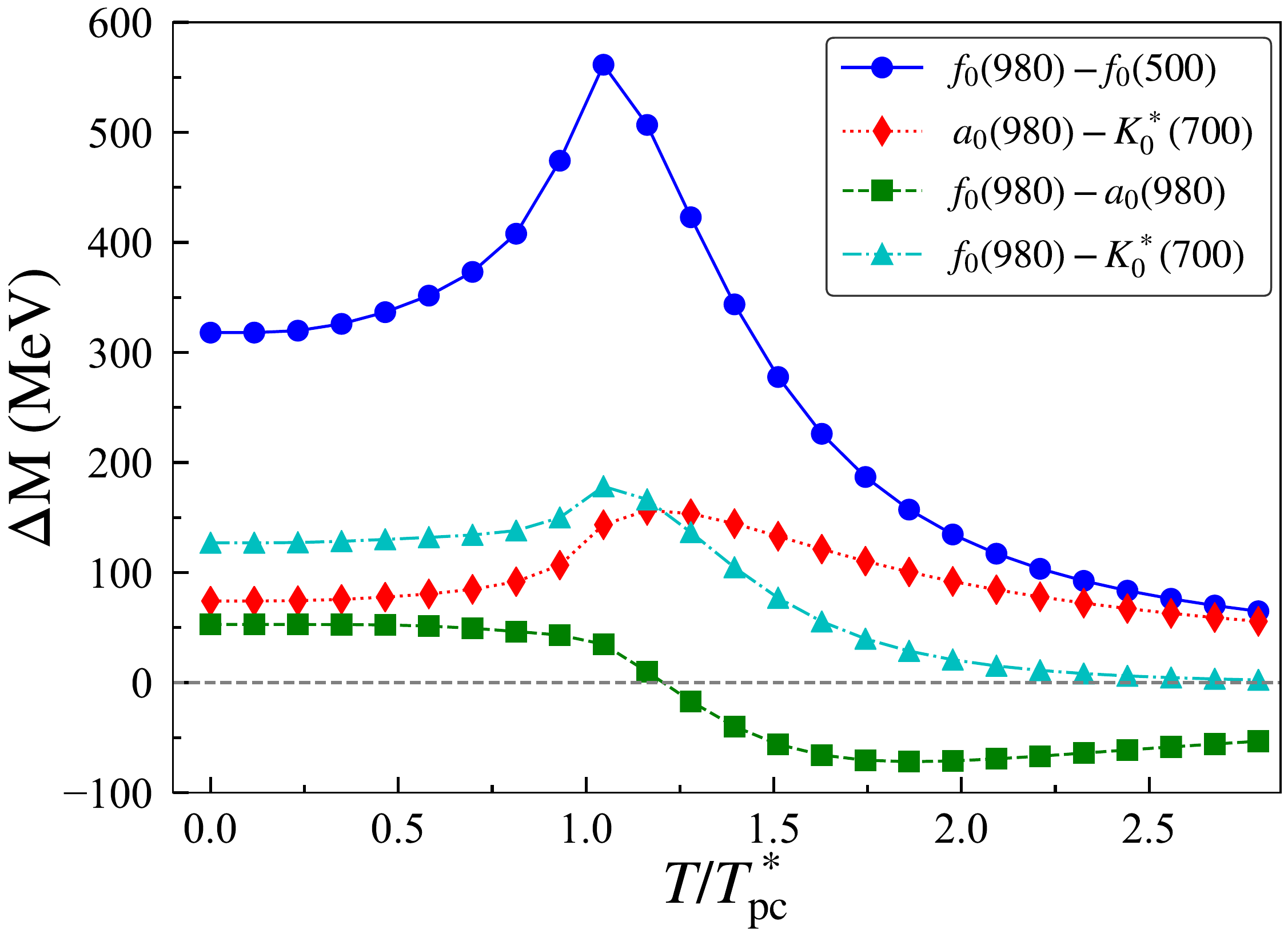}
    \subfigure{(a)}
  \end{center}
 \end{minipage}%
 \begin{minipage}{0.5\hsize}
  \begin{center}
   \includegraphics[width=7.4cm]{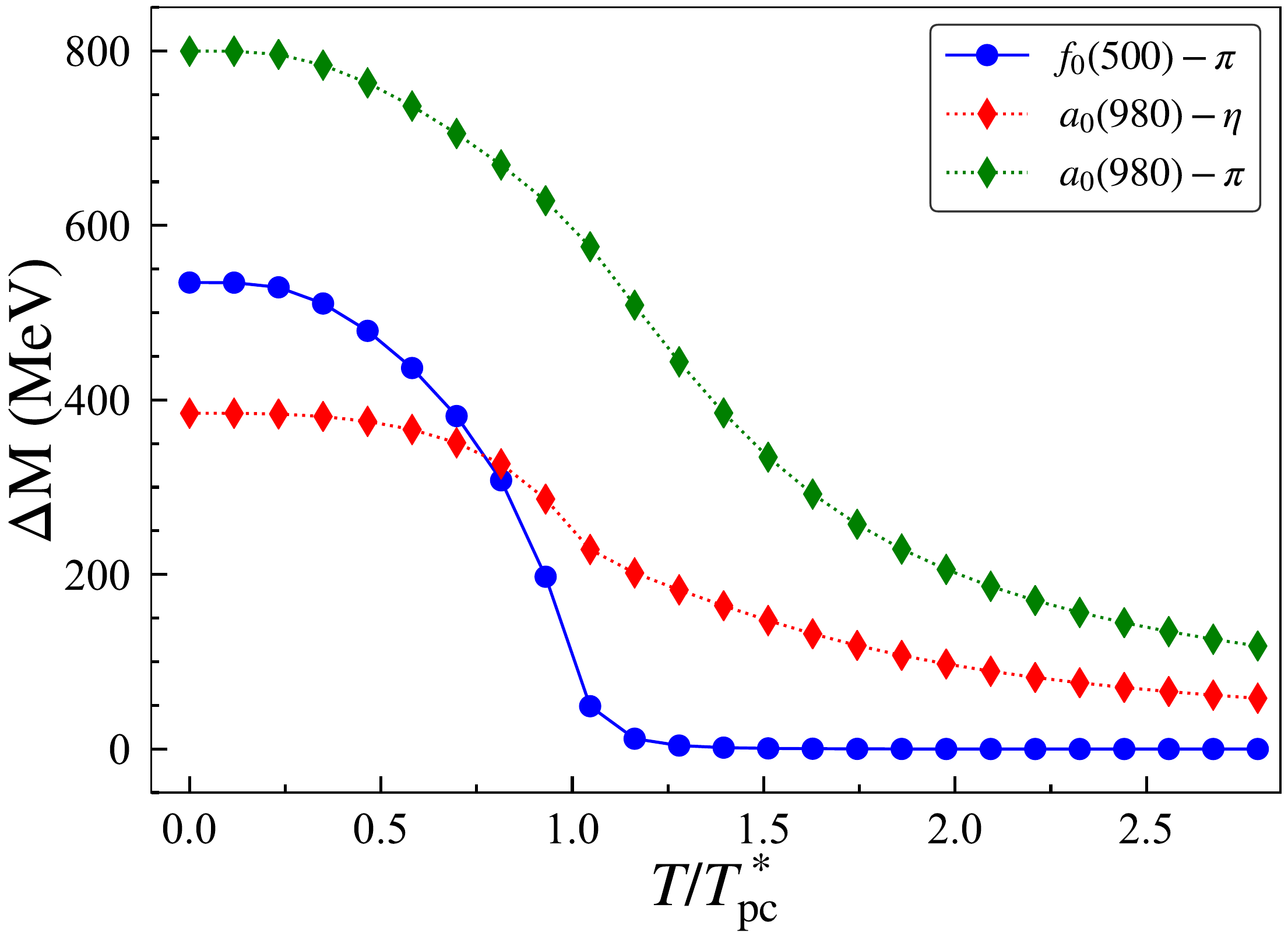}
    \subfigure{(b)}
  \end{center}
 \end{minipage}
 \end{tabular}
 \caption{ 
Evolution of the mass differences in temperature 
for (a) scalar mesons, (b) chiral and $U(1)_A$ partners.
}  
\label{mass_dif}
\end{figure}

As was clarified in~\cite{Kuroda:2019jzm}, 
we first recall that 
the axial-anomaly induced flavor breaking serves as an important source to realize the inverse mass hierarchy for scalar mesons below $1$ GeV, $m[f_0(980)]>m[a_0(980)]>m[K_0^*(700)]>m[f_0(500)]$. 
To examine the thermal effect on the inverse mass hierarchy, we plot the mass differences among scalar mesons 
in the left panel of Fig.~\ref{mass_dif}. 
We can see that the mass difference between $m[a_0(980)]$ and $m[K_0^*(700)]$ keeps the inverse mass hierarchy, $m[a_0(980)]>m[K_0^*(700)]$, 
even at high temperatures. 
This implies that the axial-anomaly induced-flavor breaking survives even at high temperatures, above 
the pseudo-critical temperature, 
and continues to keep the mass difference as 
in vacuum. 
This surviving flavor breaking in the mass difference 
$m[a_0(980)]-m[K_0^*(700)]$ might be a crucial signal 
to indirectly detect the anomaly-induced flavor breaking 
at hot QCD, maybe, at lattice QCD in the future.

Furthermore, in the left panel of Fig.~\ref{mass_dif} one can find that the order between $m[f_0(980)]$ and $m[a_0(980)]$ flips around $T_{pc}^*$:  
\begin{align} 
 m[a_0(980)]>m[f_0(980)]>m[K_0^*(700)]>m[f_0(500)] 
\qquad \textrm{at $T/T_{pc}^* \sim 1$} 
\label{ms-critical}
\end{align} 
This hierarchy flipping happens due to 
nonpertubative thermal-meson loop corrections, 
which have triggered a new flavor-breaking structure 
in the scalar meson spectra, and it is worth exploring 
in lattice simulations in the future.


In the right panel of Fig.~\ref{mass_dif}, 
we show the mass differences for the chiral partners and the $U(1)_A$ partners.
As the temperature increases, 
the mass difference of the chiral partner between $f_0(500)$ and pion becomes smaller.  
Above the $T^*_{pc}$, the $f_0(500)$ meson becomes approximately degenerate with pion. 
Recall back that the $f_0(500)$ meson is almost an  $l\bar l$ state scalar. 
Therefore, the degeneracy between $f_0(500)$ and pion may imply the restoration signal for the light quark-chiral symmetry. 
On the other hand,  the $\eta$ meson is identified as an  $s\bar s$ pseudoscalar state, where the strange quark condensate still has a sufficiently large value compared to the $\langle\bar ll \rangle$ as displayed in Fig.~\ref{RGI_qcon}. 
The consequence of the surviving $\langle\bar ss \rangle$ has been also been reflected in the mass difference between $a_0(980)$ and $\eta$: 
They tend to most slowly merge among the chiral partners 
and still keep the sizable size ($\gtrsim 100$ MeV) 
even when the $f_0(500)$ and pion get almost degenerate 
at around $T_{pc}^*$. 
This is due to the sizable strange quark contribution to both 
$a_0(980)$ (from the $k$ term and nonperturbative thermal loops) 
and $\eta$ (from nonperturbative thermal loops). 
Thus, the flavor breaking between the light- and strange-quark condensates  
is certainly reflected in the mass difference of the chiral partners. 

Note also 
(from the green curve in the right panel of Fig.~\ref{ploteffmass}) that 
the mass difference of the $U(1)_A$ partner between $a_0(980)$ and pion 
does not become degenerate even at high temperatures enough. 
This implies that the $U(1)_A$ symmetry is still broken even above $T_{pc}^*$, in accordance with the surviving 
flavor breaking induced from the axial anomaly, 
observed in the scalar meson mass spectra.

The right panel of Fig.~\ref{mass_dif} also shows 
the approximate restoration of the $O(4)\, (\simeq SU(2)_L\times SU(2)_R)$ symmetry at high temperatures, which is signaled by the degeneracy of the chiral partner, $f_0(500)$ and pion. 
Of interest is that the $U(1)_A$ symmetry detected by the degeneracy of the $U(1)_A$ partner is still broken even after the (approximate) $O(4)$ restoration. 
\footnote{In Ref.~\cite{GomezNicola:2017bhm} based on 
the Ward Identities analysis, 
it is argued that the $O(4)$ symmetry and the $U(1)_A$ symmetry are simultaneously restored by referring to  
the vanishing condition for the topological susceptibility $\chi_{\rm top}$,    
where 
the flavor breaking effect is not taken into account. 
So, we may suspect that 
the contribution of the flavor breaking would cause this discrepancy between our work and the Ward Identities analysis~\cite{GomezNicola:2017bhm} 
with the flavor symmetry assumed. }
This tendency is in agreement with the lattice observation with the physical pion mass ~\cite{Bhattacharya:2014ara}.




\subsection{Topological susceptibility}

The generic formula of $\chi_{\rm top}$ in Eq.~(\ref{linear_chi}) is still available even at finite temperatures.
Using the quark condensates in Eq.~(\ref{quark_con}), we evaluate the temperature dependence of the topological susceptibility in Eq.~(\ref{linear_chi}). 
What we report here will be summary of the main result in 
the literature~\cite{Kawaguchi:2020kdl}, with some discussions compensated. 

Some nonperturbative analyses on the 
QCD topological susceptibility at finite temperatures have so far 
been done based on chiral effective models~\cite{Fukushima:2001hr,Jiang:2012wm,Jiang:2015xqz,Lu:2018ukl}. 
However, no discussion on the correlation with 
quark condensates was made because their topological 
susceptibilities do not hold the flavor singlet form 
as in Eq.(\ref{linear_chi}) (see also Eq.(\ref{singlet_con}), for 
the flavor singlet condition), hence it seems 
to have been impossible to find the nonperturvative flavor breaking as addressed in the present paper.

\begin{figure}[H]
  \begin{center}
   \includegraphics[width=10cm]{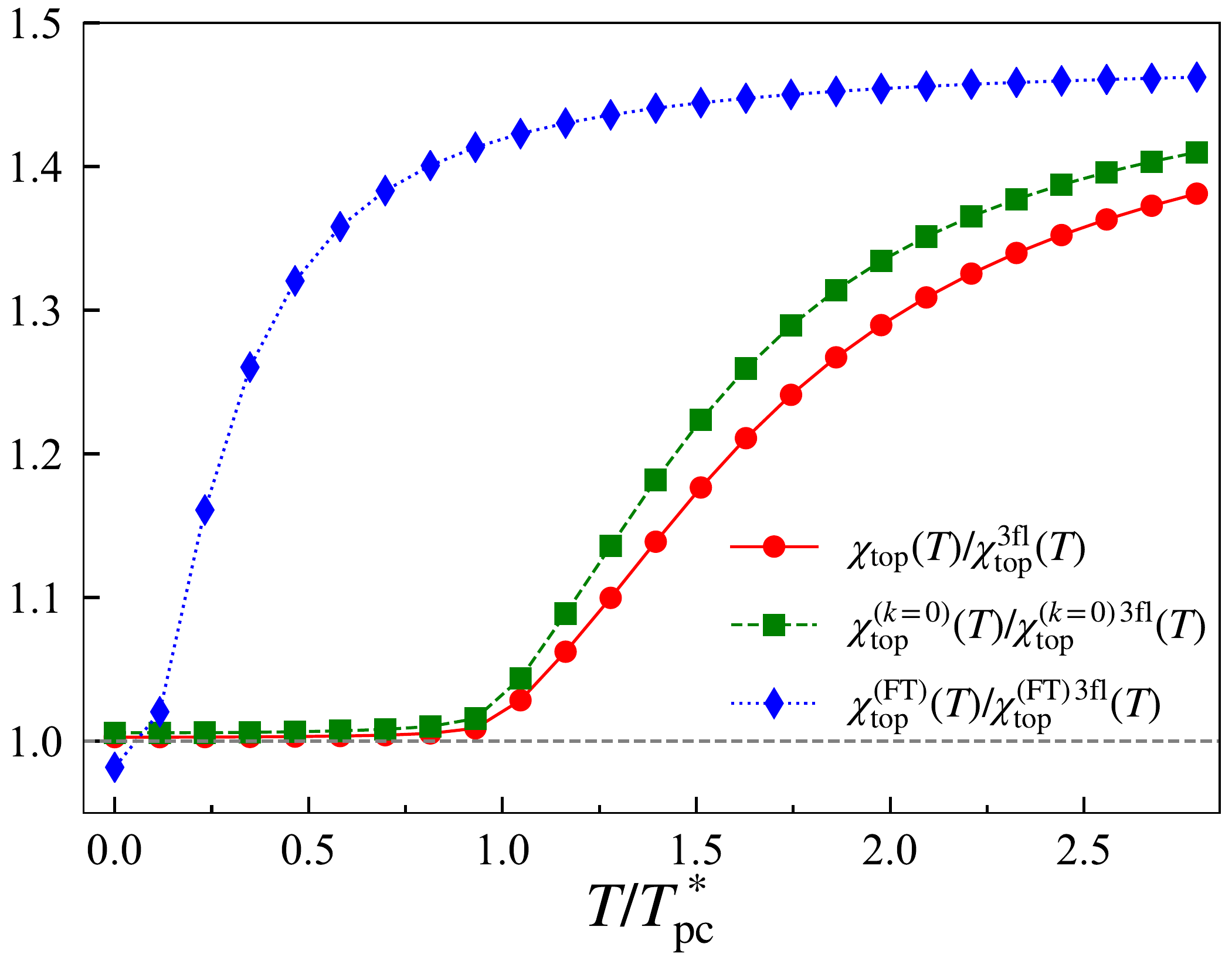}
  \end{center} 
 \caption{ 
The temperature dependence of the $\chi_{\rm top}(T)/\chi_{\rm top}^{3{\rm fl}}(T)$ compared with the one without the the anomaly-induced flavor breaking, the $k$-term, and the case in the free-quark theory, $\chi_{\rm top}^{({\rm FT})}(T)/\chi_{\rm top}^{({\rm FT})\,3{\rm fl}}(T)$. 
}  
 \label{chitopN3fl}
\end{figure}

 
As was shown in Fig.~\ref{RGI_qcon}, at high temperatures, the flavor symmetry between the light quark condensate and the strange quark condensate is drastically broken. 
Therefore, according to Eq.(\ref{linear_chi}), 
a similarly significant flavor breaking is expected 
in the topological susceptibility. 
Indeed, that takes place, see 
Fig.~\ref{chitopN3fl}. 
The 
flavor breaking in
$\chi_{\rm top}$ gets larger and larger after 
reaches the temperature around the chiral crossover, $T_{pc}^* \simeq 215$ MeV, 
precisely in the same way as the ratio of 
quark condensates ($m_s \langle \bar{s}s \rangle(T)/m_l \langle \bar{l} l\rangle(T) \gg m_s/m_l$) 
follow in Fig.~\ref{RGI_qcon}.

To extract the strange quark contribution in the topological susceptibility, in Fig.~\ref{chitopN3fl}
we show the temperature dependence of $\chi_{\rm top}(T)$ normalized to the one in the three-flavor universal limit, $\chi_{\rm top}^{3{\rm fl}}=\left(\frac{2}{m_l} + \frac{1}{m_s} \right)^{-1} \langle \bar ll \rangle(T)$.
We see that due to the sizable strange quark condensate as seen from Fig.~\ref{RGI_qcon},
the strange quark contribution to
the topological susceptibility is rapidly enhanced from $T\simeq T_{pc}^*$.
This enhancement 
is essentially driven by the sizable strange quark condensates acting as a catalyzer for the $U(1)_A$ breaking above the pseudo-critical temperature.
Eventually, after arriving at the high temperature regions where $m_s \langle \bar ss\rangle(T)/m_l \langle \bar ll\rangle(T)$ reaches the trivial-flavor breaking value $(m_s/m_l)^2$
,  
the $\chi_{\rm top}(T)/\chi_{\rm top}^{3{\rm fl}}(T)$ merges with the quark-FT regime, to asymptotically converge to $\chi_{\rm top}(T)/\chi_{\rm top}^{3{\rm fl}}(T)\simeq 3/(2+m_l/m_s)\simeq1.5$. 
Thus, 
the topological susceptibility gets the nonperturbative flavor breaking at 
around the chiral crossover criticality, which is manifestly different from 
the trivial-flavor breaking as seen in the quark-FT. 

It is remarkable that 
the catalysis of the $U(1)_A$ breaking by the nonperturbative strange quark condensate may account for the tension in the effective restoration of the $U(1)_A$ symmetry
currently observed on lattices with the two-flavor~\cite{Cossu:2013uua,Tomiya:2016jwr,Suzuki:2019vzy,Suzuki:2020rla}
and the 2+1 flavor~\cite{Bazavov:2012qja,Buchoff:2013nra,Ding:2019prx} near the chiral limit.

In the figure, also has been displayed comparison with the case without the anomaly-induced flavor breaking, the $k$-term. We see that the $k$-term plays a role of the destructive interference for the strange quark contribution to the $\chi_{\rm top}$, as was seen in Fig.~\ref{RGI_qcon}. 


\begin{figure}[H]
\begin{center}
   \includegraphics[width=7.0cm]{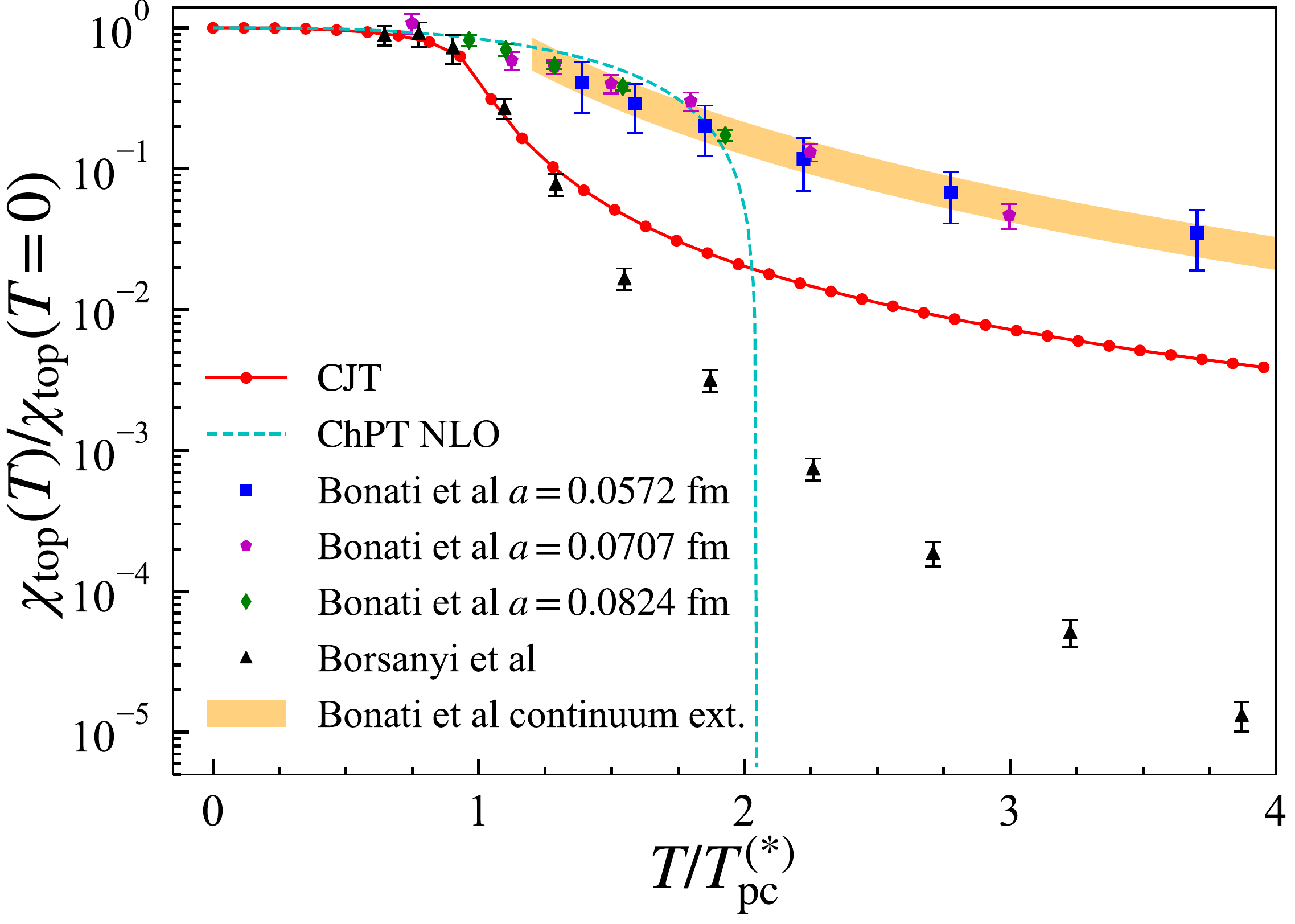}
    \subfigure{(a)}
   \includegraphics[width=7.0cm]{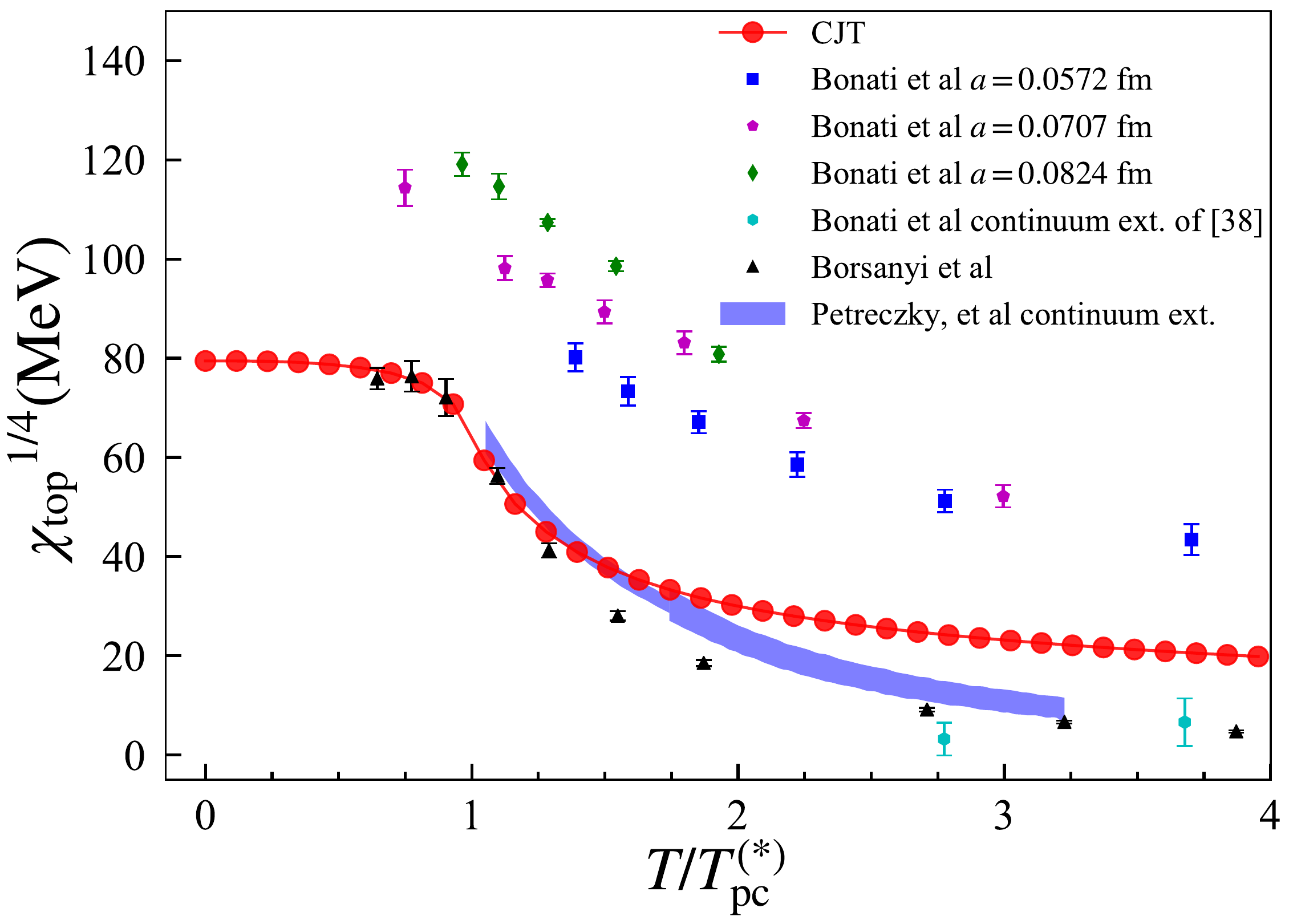}
    \subfigure{(b)}
  \end{center}
 \caption{
(a):
The comparison of the $\chi_{\rm top}(T)/\chi_{\rm top}(T=0)$ (labeled as ``CJT" in the plot) with the ChPT prediction up to the next-to-leading order (NLO)  (one-loop)~\cite{diCortona:2015ldu} and the lattice QCD data~\cite{Bonati:2015vqz,Borsanyi:2016ksw}. 
We have taken $T_{\rm pc}^*=215$ MeV for the CJT result, and 
$T_{\rm pc}=155$ MeV for the lattice simulations and the ChPT prediction.
The band corresponds to the continuum extrapolation of the lattice QCD data~\cite{Bonati:2015vqz}, which is estimated by the function
$\chi(a,T)/\chi(a,T=0)=D_0(1+D_1 a^2)(T/T_c)^{D_2}$ with 
$D_0=1.17$ $D_1=0$, $D_2=-2.71$ and $T_c=155$ MeV.
(b):
The unnormalized topological susceptibility, 
${\chi_{\rm top}}^{1/4}\,({\rm MeV})$, versus temperature, 
in comparison with 
the continuum extrapolated results 
of the recent lattice simulations.
The continuum extrapolated values of Bonati, et al 
are evaluated up to $O(a^4)$ corrections, where $a$ denotes 
the lattice scaling~\cite{Bonati:2018blm}.
The band corresponds to the continuum extrapolation of the lattice QCD data~\cite{Petreczky:2016vrs}.
}
 \label{chitopvslat}  
\end{figure}

The panel of (a) in Fig.~\ref{chitopvslat} shows the $\chi_{\rm top}$ normalized 
to the vacuum value, 
in comparison with the ChPT prediction up to the next-to-leading order (NLO)~\cite{diCortona:2015ldu} and the recent lattice data with 2 + 1 (+1) flavors having a physical pion mass and the continuum limit being taken~\cite{Bonati:2015vqz,Borsanyi:2016ksw}
 \footnote{
Note that in Ref.~\cite{Bonati:2015vqz},
data have been taken only for two different lattice spacings. 
So,continuum extrapolation in Ref.~\cite{Bonati:2015vqz} may not reliably be  performed to give predictions. 
 }
. 
The predicted $T$ dependence (denoted by ``CJT" in the figure) is actually slower-damping, 
and is for $T/T_{pc}^* \lesssim 1.5$ in good agreement 
with the lattice QCD data~\cite{Bonati:2015vqz,Borsanyi:2016ksw}, 
 which is not realized by the ChPT. 
This would manifest the importance of nonperturbative thermal contribution including the enhanced flavor breaking by $m_s \langle \bar{s}s \rangle(T)/m_l \langle \bar{l} l\rangle(T) \gg m_s/m_l$ above $T_{pc}^*$, as depicted in Fig.~\ref{RGI_qcon}. 
Substantial deviation from the lattice data can be seen for $T/T_{pc}^* > 1.5$. 
The model-prediction curve would also deviate from the one derived from the dilute instanton gas approximation~\cite{Gross:1980br,Schafer:1996wv}, which follows the lattice data~\cite{Borsanyi:2016ksw} in whole temperature range. 
This can be understood as follows. 
The present linear-sigma model picture may break down at high $T$, where degrees of freedom of quarks (and gluons) turn to become important. 
One may estimate such a $T$ as $T \sim m_q^{\rm constituent} \sim 330$ MeV, where the latter denotes mass of constituent up and down quarks (given roughly by one-third of proton mass). 
This $T$ corresponds to $T/T_{pc}^* \sim 1.5$. 
Therefore, the present model prediction may be reliable only 
up until $T/T_{pc}^* \sim 1.5$. 
Note, however, that 
it is sufficient for $T/T_{pc}^* \lesssim 1.5$ to explore the nonperturbative flavor violation in the $\chi_{\rm top}$, which 
starts to be eminent at around the chiral crossover $T/T_{pc}^* \gtrsim 1$. 

Below the chiral crossover, $T < T_{ pc}^*$, the present model perfectly 
fits with the ChPT prediction and lattice data. 
This implies that the chiral symmetry for light quarks is 
essential in the low-temperature region, as discussed in the literature~\cite{diCortona:2015ldu}
. 
Note that the ChPT-governed domain is intact even if one includes  the next-to-next-to-leading order (NNLO) correction in the ChPT analysis~\cite{Nicola:2019ohb}. 
Beyond the ChPT-governed domain, above $T_{pc}^*$ the strange quark condensate 
would serve as an important source to develop the topological susceptibility, as the consequence of the nonperturbative flavor breaking. 

In the panel (b) of Fig.~\ref{chitopvslat}
 \footnote{
The result from Ref.~\cite{Petreczky:2016vrs} is included into the panel (b) of Fig.~\ref{chitopvslat}, in which a Nambu-Goldstone pion mass is 160 MeV, 
somewhat larger than the physical pion mass.   
The continuum extrapolation data on the topological susceptibility 
are shown to be close to the subsequent observation of another
lattice simulation group~\cite{Borsanyi:2016ksw}.
 %
 }
, 
we also plot the temperature dependence of an unnormalized susceptibility  $\chi_{\rm top}^{1/4}({\rm MeV})$, where  
comparison with some recent lattice simulations with the extrapolation to the continuum limit is available~\cite{Borsanyi:2016ksw, Petreczky:2016vrs,Bonati:2018blm}. 
We see that at around the chiral crossover $T/T_{pc}^*\sim 1 - 1.5$ 
the CJT result is in good agreement with 
those continuum extrapolated data.  
Although 
qualitatively having agreement,  
for $T/T_{pc^*} \gtrsim 1.5$  
the CJT result tends to predict a somewhat larger $\chi_{\rm top}^{1/4}({\rm MeV})$. 
This discrepancy gets more larger as $T/T_{pc}^*$ 
gets larger and larger. 
This would be subject to the model-systematic error (about $30$\%) 
and validity of the linear sigma model description, as noted above.

To reconcile the gap observed when $T/T_{pc}^* \gtrsim 1.5$, one 
may note that 
model parameters in the chiral effective model can 
actually have intrinsic-temperature dependence, 
which could mimic a part of nondecoupling effects from 
integrating out quarks and/or gluons. 
For instance, 
as noted in \cite{Pisarski:2019upw}, 
the instanton study predicts 
the parameter $B$ in the $U(1)_A$ anomalous part 
to have an intrinsic-temperature dependence, which is relevant to the QCD topological structure at high temperatures. Thus, such an 
intrinsic-temperature dependence might pull   
the CJT result down for $T>T_{pv}^*$, to be fullly consistent with the lattice results in the continuum limit~\cite{Borsanyi:2016ksw, Petreczky:2016vrs,Bonati:2018blm}.

\section{Summary and conclusion}

In this paper, 
we have explored quark-flavor violation effects 
at high temperatures, 
induced from nonperturbative thermal loop corrections and axial anomaly. 
Working on a three-flavor linear-sigma model 
including an axial-anomaly induced-flavor breaking term, 
we employed a nonperturbative analysis following 
the Cornwall-Jackiw-Tomboulis formalism. 
It was shown that 
the model undergoes a chiral crossover 
with a pseudo-critical temperature, 
consistently with lattice observations.  

Regarding the flavor violation, what we have found is summarized as follows:

\begin{itemize} 

\item 

Nonperturbative thermal loop corrections 
drive up-and down-quark condensates to drop 
faster than the strange quark's toward the criticality, but still keep nonzero value 
even going far above the pseudo-critical temperature. 
In particular, just above the criticality 
the flavor breaking in the quark-condensate 
ratio $m_s \langle \bar{s}s\rangle(T)/m_l \langle \bar{l}l \rangle(T)$ 
has been shown to scale much differently from a
trivial flavor violation predicted from the ideal quark gas, and is contrast to the chiral perturbation theory predicting almost flavor-symmetric quark condensates. 
This manifests the nonperturbative generation of a significant flavor breaking at the chiral criticality (Fig~\ref{RGI_qcon});

\item 

The anomaly-related flavor-breaking effect acts as a catalyzer for the chiral restoration, while it reduces 
the amount of flavor breaking in the up, down and strange condensates (Fig.~\ref{quark_con_plot}); 

\item 

The meson flavor mixing structures are drastically 
affected in passing the chiral 
crossover, due to the nonperturbative 
thermal loop effects (Fig.~\ref{mixing_plot}), 
in which the anomaly-induced flavor breaking is found to be almost irrelevant;

\item 

The meson spectroscopy gets corrected by the net nonperturbative flavor breaking effects around and above the critical temperature (Figs.~\ref{ploteffmass} and \ref{mass_dif}), where the scalar meson mass hierarchy is significantly altered by the presence of the anomaly-related flavor breaking (Eq.(\ref{ms-critical})); 

\item 

Above the pseudo-critical temperature, 
the topological susceptibility is highly enhanced 
due to the surviving strange quark condensate compared to the three-flavor univeral limit, 
which cannot be detected by the chiral perturbation 
theory (the panel of (a) in Fig.~\ref{chitopvslat}).  
The predicted scaling in temperature 
significantly deviates from the dilute instanton gas prediction. 
There the anomaly-induced flavor breaking plays a role 
of the destructive interference for the enhancement, 
as in the flavor breaking in the quark condensates 
(Fig.~\ref{chitopN3fl}).

\item

It is noteworthy that 
$U(1)_A$ breaking perceived by the topological susceptibility is catalyzed by the nonperturbative strange quark condensate, which may account for the tension in the effective restoration of the $U(1)_A$ symmetry
currently observed on lattices with two  flavors~\cite{Cossu:2013uua,Tomiya:2016jwr,Suzuki:2019vzy,Suzuki:2020rla}
and 2+1 flavors~\cite{Bazavov:2012qja,Buchoff:2013nra,Ding:2019prx} near the chiral limit.

\end{itemize}

These predictions should, in a qualitative sense, be understood as new chiral critical features in hot QCD. 
Precise comparison with lattice simulations may be done 
by taking into account a possible systematic error (e. g. considering the intrinsic-temperature dependence on the model parameters), which 
stems from modeling low-energy QCD as a linear sigma model 
description. The systematic error 
might be estimated 
by referring to the pseudo-critical temperature $T_{pc}^*$, 
which we observed to be $\simeq 215$ MeV, while 
the recent lattice simulations have reported $T_{pc} 
\simeq 155$ MeV. The discrepancy, regarded as  
the systematic error of the present analysis, is about 
30\%. 
(Furthermore, going beyond the Hartree approximation that we have worked on, our pseudo-critical temperature might also 
get close to 
the lattice observation.) 
Though having such uncertainty, 
the present analysis would certainly grab essential features 
at the hot QCD criticality, e.g., the nonperturbative flavor breaking, which would be seen as a consequence of the significant deviation in the criticality scaling 
for the topological susceptibility from the dilute instanton gas prediction, which is on almost the same trajectory 
as the data from Ref.~\cite{Borsanyi:2016ksw} in the panel (a) of Fig.~\ref{chitopvslat}. 
Comparison also with other continuum extrapolated lattice data~\cite{Borsanyi:2016ksw, Petreczky:2016vrs
} 
in the panel (b) of Fig.~\ref{chitopvslat} implies that 
around the chiral crossover ($T/T_{pc}^* \sim 1 - 1.5$), 
the CJT result is in good agreement with the continuum extrapolated 
lattice data. 
Although the CJT result is deviated from the lattice results for 
$T/T_{\rm pc} \gtrsim 1.5$,
it would sound reasonable 
to consider 
the present model to be overall on a right track the hot QCD keeps,
with the possible systematic error of 30\% for the present model 
analysis taken into account. 
Note also that as discussed in~\cite{Ding:2017giu,Bonati:2018blm}, 
the reduction of various systematic uncertainties 
in the lattice $\chi_{\rm top}$ measurements   
is still underway. 
 In future we expect 
a more precise measurement of $\chi_{\rm top}$
with the errors in the continuum band shrank even further. 

The nonperturbative flavor breaking, especially, the significance of the strange quark condensate, in the 
topological susceptibility at around the pseudo critical temperature would give an  
impact on applications to QCD axion cosmological 
models. 
The epoch, in which QCD axion starts to roll  
and oscillate as well as the position in the potential,  
is crucial to estimate the relic abundance of the axion as a dark matter today. 
This is subject to the temperature (or time) dependence 
of the $\chi_{\rm top}$ (which corresponds to the potential height) 
at around and/or above the chiral crossover boundary. 
As recently investigated in~\cite{Kim:2018knr}, 
below and above this crossover boundary,  
the two-flavor chiral perturbation theory and dilute instanton gas descriptions have separately been 
applied in evaluating the $\chi_{\rm top}$   
(with imposing a continuity condition between two domains through 
macroscopic thermodynamics quantities). 
Our present work would improve or refine this existing approach 
by including the strange quark contribution with use of the linear 
sigma model description, instead of the dilute instanton gas, 
at around and/or above the chiral crossover. 
This would provide 
a complementary evaluation of the QCD axion cosmology, with higher reliability. 
Detailed study is to be pursued in another publication.

Similar significance of the nonperturbative flavor breaking 
can also be observed in other chiral effective models, like 
Nambu-Jona-Lasinio models with a nonperturbative formalism 
for thermal loop corrections, e.g. functional renormalization 
group method. 
This would be a cross-check on what we have addressed in this paper, and 
could give some new interpretation from a different point view 
in terms of quark degrees of freedom and renormalization group. 
It would also be interesting to work along this line 
in the future.

\section*{Acknowledgements} 

We are grateful to Massimo D’Elia, Xu-Guang Huang and Robert Pisarski 
for useful comments. 
This work was supported in part by the National Science Foundation of China (NSFC) under Grant No. 11747308 and 11975108 
and the Seeds Funding of Jilin University (S.M.).  
M.K.and A.T. thank for the hospitality of Center for Theoretical Physics and College of Physics, 
Jilin University where the present work has been partially done.  
The work of A.T. was supported by the RIKEN Special Postdoctoral Researcher program.
\appendix
\section{Reduction of $\chi_{\rm top}$ into $\chi_{\rm 5,disc}$ with the flavor-singlet condition} \label{chi5disc}
The topological susceptibility with 
the flavor singlet condition (Eq.~(\ref{top_QCD}))
can be rewritten by the disconnected pseudoscalar susceptibility $\chi_{5,{\rm disc}}$.
To see it, 
we first show the correlation between the quark condensates and the pseudoscalar susceptibilities, 
through the Ward identities derived in QCD \cite{Nicola:2016jlj,GomezNicola:2017bhm}.

We begin with the generic expression for the expectation value of 
an operator ${\cal O}(x_1)$ evaluated by the QCD generating functional,
\begin{eqnarray}
\left\langle{\cal O}(x_1)\right\rangle&=& 
Z^{-1}_{\rm QCD}
\int [\Pi_f dq_f d\bar q_f] [dA]
{\cal O} (x_1)
\exp\left(-\int_T d^4x 
{\cal L}_{\rm QCD}
\right). \label{OVEV}
\end{eqnarray}
Then we consider the chiral $SU(3)$ transformation for this expectation value. 
Under the $SU(3)_A$ flavor symmetry, the quark field transforms 
with the infinitesimal rotation angle $\alpha_A^a(x)$ as
\begin{eqnarray}
q(x)&\to&q(x) +\left(
i\alpha_A^a(x)
\frac{\lambda_a}{2}\gamma_5 q(x)
\right)\;\;\;({\rm for}\;\;a=1,\cdots8).
\end{eqnarray}
So, the $SU(3)_A$ transformation of $\langle {\cal O}(x_1) \rangle$ in Eq.(\ref{OVEV}) 
goes like  
\begin{eqnarray}
&&
\int_T d^4x
\left\langle \frac{\partial {\cal O}'(x_1)}{\partial \alpha_A^a(x) }
\right\rangle
+
\int_T d^4x
\left\langle i{\cal O}(x_1)
\partial_\mu^{(x)}j_5^{a\mu}(x)
\right\rangle
+\int_T d^4x\left\langle {\cal O}(x_1)
\left(
\bar q  \left\{ \frac{\lambda_a}{2}, {\cal M} \right\} \gamma_5 q(x)
\right)
\right\rangle
 =0,
 \label{WI_axial}
\end{eqnarray}
where integration over space-time ($T$) has been taken, and 
${\cal O}'(x_1)$ is the transformed operator 
and $j_5^{a\mu}$ represents the axial current,
$ j_5^{a\mu}(x)
=
\bar q \gamma^\mu \gamma_5  \frac{\lambda_a}{2}q(x)$.
Equation (\ref{WI_axial}) is the Ward-identity for $SU(3)$ chiral symmetry.

The second term in the left-hand side of Eq.~(\ref{WI_axial}) is evaluated as 
\begin{eqnarray}
&&\int_T d^4x
\left\langle i{\cal O}(x_1)
\partial_\mu^{(x)}j_5^{a\mu}(x)
\right\rangle\nonumber\\
&=&
\int_T d^4x\partial_\mu^{(x)}
\left\langle0|T  i{\cal O}(x_1)
j_5^{a\mu}(x)
|0\right\rangle\nonumber\\
&=&
\begin{cases}
\left\langle0\left|
[iQ_5^a(x_1^0),{\cal O}(x_1)]
\right|0
\right\rangle=
-\int_T d^4x
\left\langle \frac{\partial {\cal O}'(x_1)}{\partial \alpha_A^a(x) }
\right\rangle
&{\rm (Chiral\; limit)}
\\
0&{\rm (Explicit\;chiral\;symmetry\;breaking)}, 
\end{cases}
\end{eqnarray}
where $Q_5^a$ is the $SU(3)_A$ charge defined as $Q_5^a(x^0) \equiv \int d^3 x j^{a\mu=0}_5(x)$. 
In the chiral limit this term 
does not drop even though it is just a surface term,  
due to 
emergence of the massless NG (Nambu-Goldstone) bosons
associated with the spontaneous chiral symmetry breaking, which 
survive in the infinitely long-distance $(|x_1 - x|\to \infty)$.
In contrast, 
the explicit chiral symmetry breaking gives NG bosons nonzero masses, so that 
$\left\langle0|T  i{\cal O}(x_1)
j_5^{a\mu}(x) |0\right\rangle$ goes to zero when $|x_1-x|\to \infty$. 

Now we choose the ${\cal O}(x_1)$ as 
the pseudoscalar operator, 
\begin{eqnarray}
{\cal O}^b(x_1)=iq \gamma_5\frac{\lambda^b}{2}q(x_1),\;\;\;
({\rm for}\;\;b=0,1,\cdots,8).
\end{eqnarray} 
Then, 
for the $\pi$-channel corresponding to ${\cal O}^{b=1,2,3}(x_1)$ with $\alpha_A^{a=1,2,3}(x)$,  
the Ward identity in Eq.(\ref{WI_axial}) for this operator 
reads 
\begin{eqnarray}
\langle
\bar q_lq_l(x_1)
\rangle
=
-m_l
\chi_\pi(x_1), 
\label{pi_WI}
\end{eqnarray}
where $q_l$ is the light quark field, 
$q_l=(u,d,0)$, and $\chi_\pi$, called the pseudoscalar susceptibility,  
is defined as  
\begin{eqnarray}
\chi_\pi(x_1)
=
\int_T d^4x\left\langle 
\left( i\bar q_l \gamma_5q_l(x_1)\right)
 \left(i
 \bar q_l\gamma_5q_l(x)
 \right) \right\rangle_{\rm con},
\end{eqnarray}
with $\langle\cdots\rangle_{\rm con}$ representing 
the connected part of the correlation function. 
Thus, one finds that the light quark condensate is connected with the pseudoscalar susceptibility $\chi_\pi$ \cite{Nicola:2016jlj}.

As for the $\eta$-$\eta'$ channel,   
corresponding to ${\cal O}^{b=0}(x_1)$ with $\alpha_A^{a=8}(x)$
and ${\cal O}^{b=8}(x_1)$ with $\alpha_A^{a=8}(x)$,
one can also find the correlations between the quark condensates and the pseudoscalar susceptibilities 
from the associated Ward identities~\cite{Nicola:2016jlj},
\begin{eqnarray}
\langle
\bar q_lq_l
\rangle
-2
\langle\bar ss
\rangle
&=&
-m_l(\chi_P^{ll}+\chi_P^{ls})+2m_s(\chi_P^{ss}+\chi_{P}^{ls}), 
\nonumber\\
\langle
\bar q_l q_l
\rangle
+4
\langle\bar ss\rangle
&=&
-m_l(\chi_P^{ll}-2\chi_P^{ls})-2m_s(2\chi_P^{ss}-\chi_{P}^{ls}), 
\label{eta_sector_WI}
\end{eqnarray}
where 
$\chi_{P}^{ll} $, $\chi_{P}^{ss}$ and $\chi_{P}^{ls}$ 
respectively are defined as  
\begin{eqnarray}
\chi_{P}^{ll} &=&
 \int_T d^4x
\left\langle
\left(
 i\bar q_l \gamma_5 q_l(x_1)\right)
 \left(
 i\bar q_l\gamma_5q_l(x)
 \right)
\right\rangle, 
\nonumber\\
\chi_{P}^{ss}&=&
 \int_T d^4x\left\langle
\left( i\bar s \gamma_5s(x_1)\right)
 \left(i
 \bar s\gamma_5s(x)
 \right) \right\rangle, 
 \nonumber\\
\chi_{P}^{ls}&=&
\int_T d^4x\left\langle
\left( i\bar q_l \gamma_5q_l(x_1)\right)
 \left(i
 \bar s\gamma_5s(x)
 \right) \right\rangle. 
\end{eqnarray}
From the correlation relations in Eq.~(\ref{eta_sector_WI}),
the light quark condensate and the strange quark condensate can be read as
\begin{eqnarray}
\langle\bar q_lq_l \rangle&=&
-m_l\chi_P^{ll}+2m_s\chi_P^{ls},\nonumber\\
\langle\bar ss\rangle&=&
-m_s\chi_P^{ss}+\frac{m_l}{2}\chi_P^{ls}.
\label{cond_sus}
\end{eqnarray}
In addition,
by using Eqs.~(\ref{pi_WI}) and (\ref{eta_sector_WI}),
$\chi_P^{ls}$ can be written as the disconnected pseudoscalar susceptibility~\cite{GomezNicola:2017bhm}:    
\begin{eqnarray}
\chi_P^{ls}
&=&
\frac{m_l}{2m_s}
(\chi_{P}^{ll}-\chi_{\pi})
\nonumber\\
&=&
2\frac{m_l}{m_s}\chi_{5,{\rm disc}}. 
\label{pseudo_disc}
\end{eqnarray}

Using Eqs.~(\ref{cond_sus}) and (\ref{pseudo_disc}), 
one can now find that 
the topological susceptibility in Eq.~(\ref{top_QCD}) is 
reduced to 
the disconnected pseudoscalar susceptibility: 
\begin{eqnarray} 
\chi_{\rm top}
&=&
\left(\frac{\langle \bar q_lq_l \rangle(T)}{m_l}+\frac{\langle \bar ss \rangle(T)}{m_s}
\right)\bar m^2
+
\bar m^2\left(
\chi_P^{ll}+2\chi_{P}^{ls}+\chi_{P}^{ss}
\right)
\nonumber\\
&=&
\frac{1}{2}m_l m_s\chi_P^{ls}
\nonumber\\
&=&
m_l^2\chi_{5,{\rm disc}}. 
\end{eqnarray}
Thus 
the topological susceptibility in Eq.~(\ref{top_QCD})
exactly reproduces the well-known formula 
described by $\chi_{5,{\rm disc}}$
 in the 2+1-flavor QCD.
Note that in this derivation 
we have not assume the strange quark to be heavy enough ($m_s\to \infty$), 
nor make the assumption of a nearly-exact chiral-symmetry for up and down quarks where the chiral condensate is estimated to be negligible.

\section{Quark condensate in free theory}
In the free theory of quark fields at finite temperature, the quark condensate is given as  
\begin{eqnarray}
\langle\bar q q \rangle_{\rm FT}
&=&
-i4m_q\int\frac{d^4 p}{(2\pi)^4}\frac{1}{p^2-m^2+i\epsilon}
+\frac{4m_q T^2}{2\pi^2} I(m_q)
\label{freecon}
\end{eqnarray}
where
\begin{eqnarray}
I(m_q)=\int_0^\infty dx\frac{x^2}{\sqrt{x^2+m_q^2/T^2}}\frac{1}{ 1+\exp(\sqrt{x^2+m_q^2/T^2})}.
\end{eqnarray}
For $ T \gg m_q$, the second term of Eq.~(\ref{freecon}) goes like 
\begin{eqnarray}
\frac{4m_q T^2}{2\pi^2}\int_0^\infty dx\frac{x}{ 1+e^x}.
\end{eqnarray}

Focusing only on 
the thermal 
quark loop, 
the ratio of 
the strange quark condensate to the light quark's 
is evaluated as 
\begin{eqnarray}
\frac{\langle\bar  ss \rangle_{\rm FT}(T)}{\langle\bar l l \rangle_{\rm FT}(T)}=\frac{m_s}{m_l}
\frac{I(m_s)}{I(m_l)},
\end{eqnarray}
Since $I(m_l)=I(m_s)$ 
for $T \gg m_q $, 
the ratio of quark condensates converges to $m_s/m_l$ as the temperature increases.






\end{document}